\documentclass[usenatbib]{mn2e}
\usepackage{cite}
\usepackage{graphicx}
\usepackage{epsfig}
\usepackage{amsmath,amssymb}
\usepackage{lscape}
\usepackage{verbatim}
\usepackage{rotating}
\usepackage{epstopdf}
\usepackage{rotating}
\usepackage{pdflscape}
\usepackage{subfigure}
\usepackage{chngcntr}
\usepackage{float}
\usepackage{placeins}
\usepackage{color}
\usepackage{url}

\def\lapprox{\lower.4ex\hbox{$\;\buildrel <\over{\scriptstyle\sim}\;$}}
\def\gapprox{\lower.4ex\hbox{$\;\buildrel >\over{\scriptstyle\sim}\;$}}

\DeclareRobustCommand{\ion}[2]{%
\relax\ifmmode
\ifx\testbx\f@series
{\mathbf{#1\,\mathsc{#2}}}\else
{\mathrm{#1\,\mathsc{#2}}}\fi
\else\textup{#1\,{\mdseries\textsc{#2}}}%
\fi}

\title[]{Searching for FUV line emission from $10^7$ K gas in massive elliptical galaxies and galaxy clusters as a tracer of turbulent velocities}
\author[Anderson and Sunyaev]{Michael E. Anderson$^{1}$\thanks{email: michevan@mpa-garching.mpg.de} and Rashid Sunyaev$^{1,2}$ \\
$^{1}$Max-Planck Institute for Astrophysics, Garching bei Muenchen, Germany\\
$^{2}$Space Research Institute (IKI),  Russian Academy of Sciences, Profsoyuznaya 84/32, Moscow 117997, Russia\\
}
\begin{document}

\maketitle

\begin{abstract}
Non-thermal pressure from turbulence and bulk flows is a fundamental ingredient in hot gaseous halos, and in the intracluster medium it will be measured through emission line kinematics with calorimeters on future X-ray spacecraft. In this paper we present a complementary method for measuring these effects, using forbidden FUV emission lines of highly ionized Iron which trace $10^7$ K gas. The brightest of these is [\ion{Fe}{xxi}] $\lambda$1354.1. We search for these lines in archival HST-COS spectra from the well-known elliptical galaxies M87 and NGC4696, which harbor large reservoirs of $10^7$ K gas. We report a 2.2$\sigma$ feature which we attribute to [\ion{Fe}{xxi}] from a filament in M87, and positive residuals in the nuclei of M87 and NGC4696, for which the 90\% upper limits on the line flux are close to the predicted fluxes based on X-ray observations. In a newer reduction of the data from the Hubble Spectroscopic Legacy Archive, these limits become tighter and the [\ion{Fe}{xxi}] feature reaches a formal significance of 5.3$\sigma$, neglecting uncertainty in fitting the continuum. Using our constraints, we perform emission measure analysis, constraining the characteristic path length and column density of the $\sim10^7$ K gas. We also examine several sightlines towards filaments or cooling flows in other galaxy clusters, for which the fraction of gas at $10^7$ K is unknown, and place upper limits on its emission measure in each case. A medium-resolution HST-COS observation of the M87 filament for $\sim$10 orbits would confirm our detection of [\ion{Fe}{xxi}] and measure its width.

\end{abstract}
 

%\large
\section{Introduction}

Turbulence and bulk flows provide non-thermal pressure in the hot gaseous halos of galaxies, galaxy groups, and galaxy clusters. On large scales, this is predicted by numerical simulations (e.g. \citealt{Norman1999}, \citealt{Frenk1999}) due to mergers and the accretion of satellites. In the centers of these systems, turbulence and bulk flows are also introduced by active galactic nucleus (AGN) feedback \citep{Churazov2001}; other mechanisms such as cosmic ray-driven instabilities \citep{Sharma2009} and galactic stirring (\citealt{Kim2007}; \citealt{Ruszkowski2010}) may also be important in the intracluster medium. 

These motions play a fundamental role in the physics of these systems. They transport energy, mass, and metals from the central nucleus into the diffuse halo, affecting the thermodynamic state of the intracluster medium and mediating the enrichment of the intergalactic medium. The non-thermal pressure from turbulence is also necessary to understand in order to estimate the masses of galaxy groups and clusters from hydrostatic equilibrium arguments. Large-scale flows are expected as well around the supermassive black holes at the centers of these systems, particularly as one approaches the Bondi radius \citep{Bondi1952}. In order to understand turbulence and bulk flows in better detail, a great deal of effort has been focused on the kinematics of the hot halo gas in galaxies, galaxy groups, and galaxy clusters, both with observations (e.g. \citealt{Xu2002}, \citealt{Churazov2004}, \citealt{Schuecker2004}, \citealt{Werner2009}, \citealt{Sanders2011}), and simulations (e.g. \citealt{Evrard1990}, \citealt{Schindler1993}, \citealt{Norman1999}, \citealt{Nagai2003}, \citealt{Sunyaev2003}, \citealt{Brueggen2005a}, \citealt{Brueggen2005b}, \citealt{Vazza2006}, \citealt{Lau2009}, \citealt{Battaglia2012}, \citealt{Gaspari2014}, \citealt{Zhuravleva2014} ). 

However, these measurements are extremely challenging and expensive. Constraints can be obtained in some cases through the kinetic SZ effect \citep{Sunyaev1980}, and observations of resonant scattering in the intracluster medium \citep{Gilfanov1987}. The more direct approach is to measure the broadening and redshift of emission lines, which trace turbulence and bulk motions respectively. For bulk flows, \citet{Inogamov2003} showed that the typical motions of bulk flows in galaxy clusters are much larger than the peculiar velocities of the clusters, making such measurements possible. Also, the thermal velocities of heavy ions in hot gas ($v_{\text{th}} = \sqrt{2 k T / M_i}$) are $\sim \sqrt{M_i/m_p}$ times smaller than the speed of sound ($c_s =  \sqrt{\gamma k T / \mu m_p}$) in the same gas. Measuring the broadening of (for example) Iron lines in hot plasmas, we can detect broadening due to turbulence because even subsonic turbulence with velocities on the level of 10\% of sound speed increases significantly the width of heavy ion lines \citep{Inogamov2003}.  To do this, one needs to resolve the emission lines, which are primarily the soft X-ray \ion{O}{viii} and Fe L-shell lines for gas around $10^7$ K, but CCD detectors on modern X-ray telescopes (R $\equiv \lambda / \Delta \lambda\sim 15-20$) lack the required spectral resolution. X-ray grating spectrographs have better resolution ($R \sim 100-1000$) and can identify important features like the Fe L shell lines, but these lines lie very close to each other and are usually blended. Using a calorimeter, the Hitomi satellite was recently able to make a high-precision measurement of the broadening in the Fe K$\alpha$ line for a single target - the Perseus cluster (Fabian et al 2016), which is the X-ray brightest cluster in the sky. In coming decades, such devices aboard next-generation spacecraft like Athena or the X-ray Surveyor will offer similar improvements in clusters hot enough to produce this line. For $10^7$ K gas, however, X-ray calorimeters will be limited to R$\sim$300-400 for the prominent X-ray emission lines.

Here we discuss a complementary approach: measuring the kinematics of far-ultraviolet emission lines. In the FUV, the Cosmic Origins Spectrograph (COS) instrument \citep{Green2012} on the Hubble Space Telescope (HST), offers much higher throughput and spectral resolution. Typically ultraviolet lines are characteristic of cooler plasma, at temperatures of $10^4-10^6$ K, but there are exceptions. Some FUV lines have been used as diagnostics of plasma at temperatures around $10^7$ K, both in Solar flare spectra (e.g. \citealt{Doschek1975}, \citealt{Cheng1979}, \citealt{Mason1986}, \citealt{Feldman1991}, \citealt{Innes2003a}, \citealt{Innes2003b},  \citealt{Doschek2010}, \citealt{Young2015}), and in the coronae of nearby late-type stars (e.g. \citealt{Maran1994}, \citealt{Linsky1998}, \citealt{Ayres2003}, \citealt{Redfield2003}, \citealt{Dupree2005}). In this paper we will discuss these lines in an extragalactic context, as tracers of the intracluster medium and accretion flows.  COS has one low-resolution grating (G140L) with $R \sim 10^3$ and two medium-resolution gratings (G130M and G160M) with $R \sim 1-2\times10^4$, all three of which we utilize in this work. We note that this method is conceptually similar to a method proposed by \citet{Sunyaev1984} to use a 3 mm hyperfine transition from Lithium-like $^{57}$Fe.

In Figure 1 we present basic diagnostic properties for six of the brightest of these lines. They are all ground-state magnetic dipole transitions of various species of highly ionized Iron. In the top panel, we compare the fractional abundance of these six species of Iron for plasma in collisional ionization equilibrium (CIE), based on the CHIANTI database v. 7.15 (\citealt{Dere1997}, \citealt{Landi2013}). In the middle panel, we present the effective cross section for absorption from these lines, scaled to the Thomson cross section ($6.65\times10^{-25}$ cm$^{-2}$). Only thermal broadening is considered, and we assume an Iron abundance relative to Hydrogen of $3.16\times10^{-5}$ \citep{Grevesse1998}. In most extragalactic cases these lines will be optically thin, and since they are forbidden lines the optical depths will generally be very low. They may be detectable in some situations, although in this paper we focus on the lines in emission. It is important to emphasize that, since these lines are forbidden transitions and optically thin, the observed line widths are not affected by most radiative transfer effects which could otherwise introduce artificial broadening to emission lines.

In the bottom panel of Figure 1 we present the emissivity of these lines, defined such that $\epsilon = L / (Z_{\text{Fe}} \int{n_en_pdV})$ where  $L$ is the line luminosity and $Z_{\text{Fe}}$ is the fractional abundance of Iron relative to Hydrogen, for which we assume the Solar value measured by \citet{Grevesse1998}. The quantity $\int{n_en_pdV}$ is the volume emission measure, which for a pencil-beam sightline differs from the pencil-beam emission measure $\int{n_en_pdl}$ by a factor of the area $A$ covered by the sightline. This emissivity is also derived from the CHIANTI database, by multiplying their G(T) function from \verb"gofnt.pro" by $4\pi$ steradians. It is computed assuming collisional ionization equilibrium and that the lines are optically thin. 

\begin{figure*}
\begin{center}
\includegraphics[width=16cm]{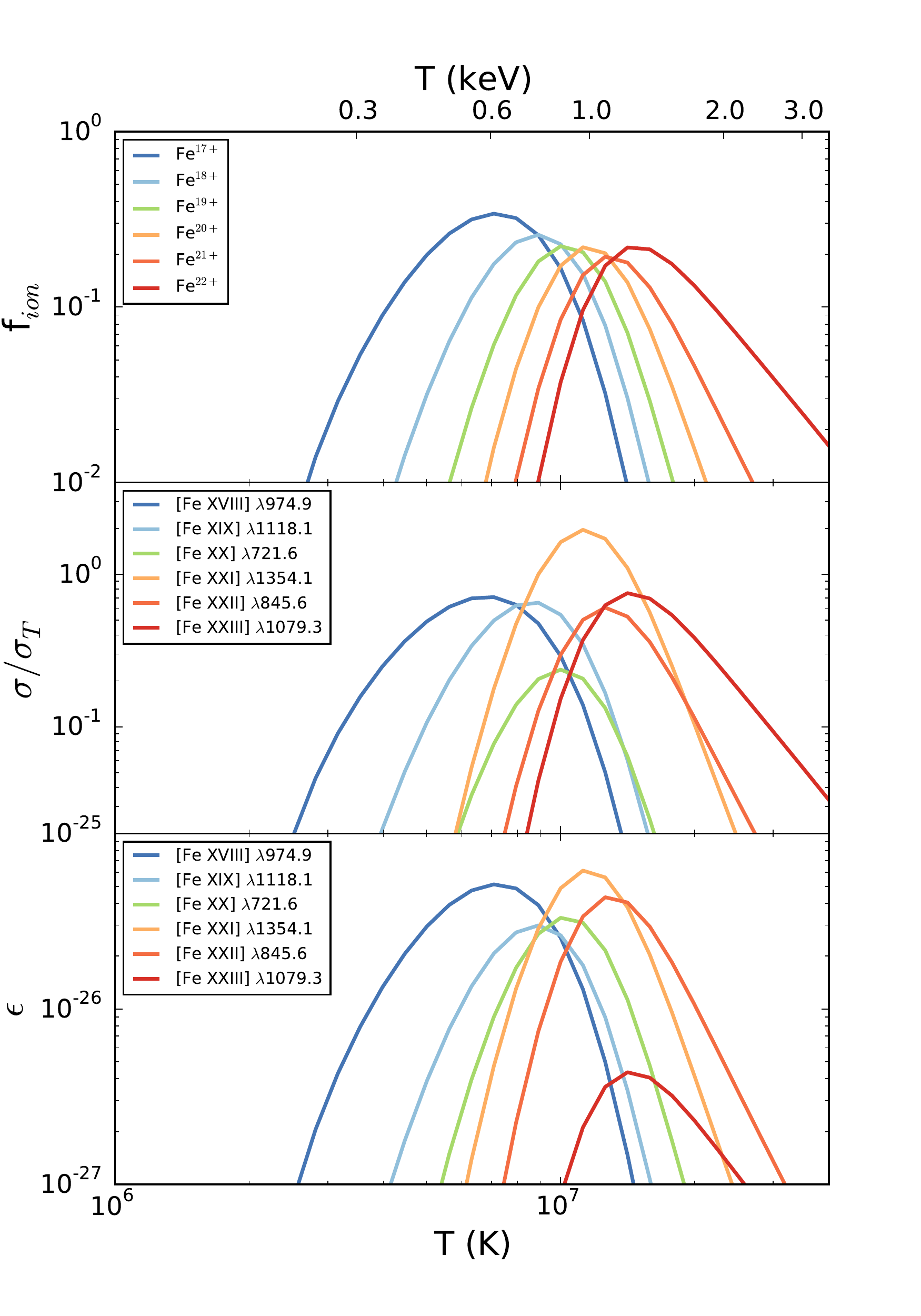}
\end{center}
\vspace{-1.0cm}
\caption{(top) Fractional abundance of six species of highly ionized Iron, for plasma in CIE. (middle) Effective absorption cross sections for the strongest FUV transition for each species, scaled to the Thomson cross section and assuming the lines are optically thin and affected only by thermal broadening. (bottom) Approximate emissivity for these transitions ($\epsilon = L / (Z_{\text{Fe}} \int{n_en_pdV})$), in units of erg cm$^3$ s$^{-1}$. Middle and bottom panels assume Solar abundance.}
\end{figure*}

In Figure 1, the brightest of these lines is [\ion{Fe}{xxi}] $\lambda$1354.08. In this paper we will focus primarily on this line, though in some cases information on this line is not available and we consider other lines. These lines trace plasma in a narrow range around $10^7$ K, with the emissivity of [\ion{Fe}{xxi}] peaking at $10^{7.05}$ K. The instrument of choice for emission-line searches is COS, since it has the highest throughput in the FUV at present. We searched the COS archive for observations of systems known to host a significant reservoir of gas around $10^7$ K. We considered M87, Cen A, NGC 4696, and HCG 92; out of these targets, M87 and NGC 4696 have already been observed by COS in the FUV. There are also observations of HCG 92, but the the sightlines target a shock front near NGC 7318b which is likely significantly cooler than $10^7$ K. We also searched for observations of H$\alpha$ filaments and/or cooling flows in galaxy clusters; the ambient media here are hotter than $10^7$ K, but at the locations of these features multiphase gas is present, which may include $10^7$ K gas. We find suitable observations for A1795, Perseus, Phoenix (SPT-CL J2344-4243), and Zw 3146. In these latter cases we lack definite predictions of massive reservoirs of $10^7$ K gas along these sightlines, but we can use upper limits to establish independent constraints on the differential emission measure. A summary of the observations we consider is given in Table 1. 

In the next several sections of this paper, we describe the results of our search.  In Section 2, we present a tentative detection of [\ion{Fe}{xxi}] and weak evidence for [\ion{Fe}{xix}] in the filaments projected 1.9 kpc from the center of M87. In sections 3 and 4 we present upper limits on [\ion{Fe}{xxi}] in the centers of M87 and NGC 4696. In Section 5 we discuss potential sources of contamination, i.e. other emission lines which could be confused with [\ion{Fe}{xxi}] and/or [\ion{Fe}{xix}] in these galaxies. In Section 6, we report limits on  [\ion{Fe}{xviii}], [\ion{Fe}{xix}], [\ion{Fe}{xxi}], and [\ion{Fe}{xxiii}] in COS observations of several additional clusters of galaxies. In Section 7, we show how the results change if the co-addition and error estimation is modified. Finally, in Section 8 we discuss prospects for future searches. In the Appendix we discuss the filtering for geocoronal emission performed for the G140L spectra towards M87.

\begin{table*}
\begin{minipage}{120mm}
\caption{Observations Examined in this Work}
\begin{tabular}{lllll}
\hline
Dataset & target & Grating & Central Wavelength  & Exposure Time \\
 &  &  & (\AA) & (s)\\
\hline 
LBK104010 & M87 filament & G140L & 1280 & 5025\\
LC7T01010 & M87 filament & G140L & 1280 & 13386 \\ 
LC7T02010 & M87 filament & G140L & 1280 & 13387 \\ 
LC8701010 & M87 nucleus & G130M & 1318 &1000\\
LC8701020 & M87 nucleus & G130M & 1309 & 1000\\
LC8701030 & M87 nucleus & G130M & 1300 & 1200\\
LC8701040 & M87 nucleus & G130M & 1291 & 1200\\
LC8702010 & NGC 4696 nucleus & G130M & 1291  & 1191 \\
LC8702020 & NGC 4696 nucleus & G130M & 1300  & 1191 \\
LC8702030 & NGC 4696 nucleus & G130M & 1318  & 1417 \\
LC8702040 & NGC 4696 nucleus & G130M & 1327  & 1417 \\
LC1J01010 & A1795 filament & G140L & 1280 & 16443 \\
LBPO01010 & Perseus filament & G140L & 1105 & 11437\\
LBPO02010 & Perseus filament & G140L & 1280 & 8310\\
LC9552010 & Phoenix nucleus & G160M & 1577 & 19752\\
LCDI03010 & Zw 3146 nucleus & G140L & 1280 & 25279\\
LCAY02010 - LCAY06040 & HCG 92 & G130M / G160M & several & 42074\\
\hline
\end{tabular}
\\
\small{List of the HST-COS observations examined in this work, along with basic properties of the observations. For HCG 92, we have combined together 20 observations of five different regions along the same shock front.}
\end{minipage}
\end{table*}

\subsection{Observing Extended Emission with COS}

It is important to emphasize that COS is not optimized for the analysis of extended emission. It employs a circular slitless aperture, so if emission is extended in the sky along the COS dispersion axis, the lines will appear artificially broadened in the resulting spectrum \citep{Green2012}. This effect is likely contributing to the abnormally large line widths seen for \ion{C}{iv} and \ion{He}{ii}, and is also plausibly occurring at the locations of [\ion{Fe}{xxi}] and [\ion{Fe}{xix}].

To estimate this effect, we fit a Gaussian to the along-dispersion vignetting profile given in Table 6 of the COS instrument science report 2013-03 \citep{Proffitt2010} and use the along-dispersion plate scale of 0.0285"/pixel to convert into velocity units at 1354.1\AA. From this analysis, we estimate $\sigma = 648$ km/s for G140L and $\sigma = 81$ km/s for G160M. We consider these values the theoretical prediction for spatial broadening of uniformly distributed [\ion{Fe}{xxi}] emission. They are a bit lower than the $\sigma \sim 110$ km/s estimated for spatially broadened lines using G130M observations of the compact blue galaxy KISSR242 \citep{France2010}, but the latter measurements include the intrinsic line widths from the galaxy as well. 

An additional effect is that \verb"calcos" pipeline is intended for the analysis of point sources, so the fluxes inferred for extended emission like [\ion{Fe}{xxi}] may be biased. To estimate this effect, we fit a fifth-order polynomial to the cross-dispersion profile from \citet{Proffitt2010} (their Table 5) and integrate this profile over the 2."5 aperture in order to estimate the expected throughput for a well-centered point source. We find that the expected throughput for uniformly distributed extended emission is about 85\% of the expected throughput for a point source. However, since the exact correction depends on the (unknown) spatial distribution of the emission, and the correction is much smaller than the current measurement uncertainties, we neglect it in the present analysis.

\section{M87 Filament}

M87 is the central dominant elliptical galaxy at the core of the Virgo cluster, and is suffused with hot X-ray emitting gas. Near the center of the galaxy, much of this hot gas is multiphase (\citealt{Belsole2001}, \citealt{Molendi2002b}, \citealt{Simionescu2008}, \citealt{Werner2010}), spanning at least a factor of six in temperature from about 3 keV to about 0.5 keV, with absorption possibly obscuring gas below this temperature \citep{Werner2013}. M87 also contains a system of filaments which are well-known H$\alpha$ emitters (\citealt{Arp1967}, \citealt{Ford1979}), and which have also recently been discovered to be bright in \ion{C}{iv} $\lambda 1549$  \citep{Sparks2009}.  A subsequent observation with the COS G140L grating shows unambiguous emission lines from both  \ion{C}{iv} and the \ion{He}{ii} \citep{Sparks2012}, indicating the presence of $10^5$ K gas in these structures. At lower temperatures, \citet{Werner2013} used the PACS instrument on Herschel to detect [\ion{C}{ii}] $\lambda$158$\mu$m  emission from M87 which seems to trace the filamentary structure as well; this is typically the dominant cooling line when Hydrogen is neutral  \citep{Dalgarno1972} and is most effective at temperatures around $10^2$ K.

\begin{figure*}
\begin{center}
\includegraphics[width=17cm]{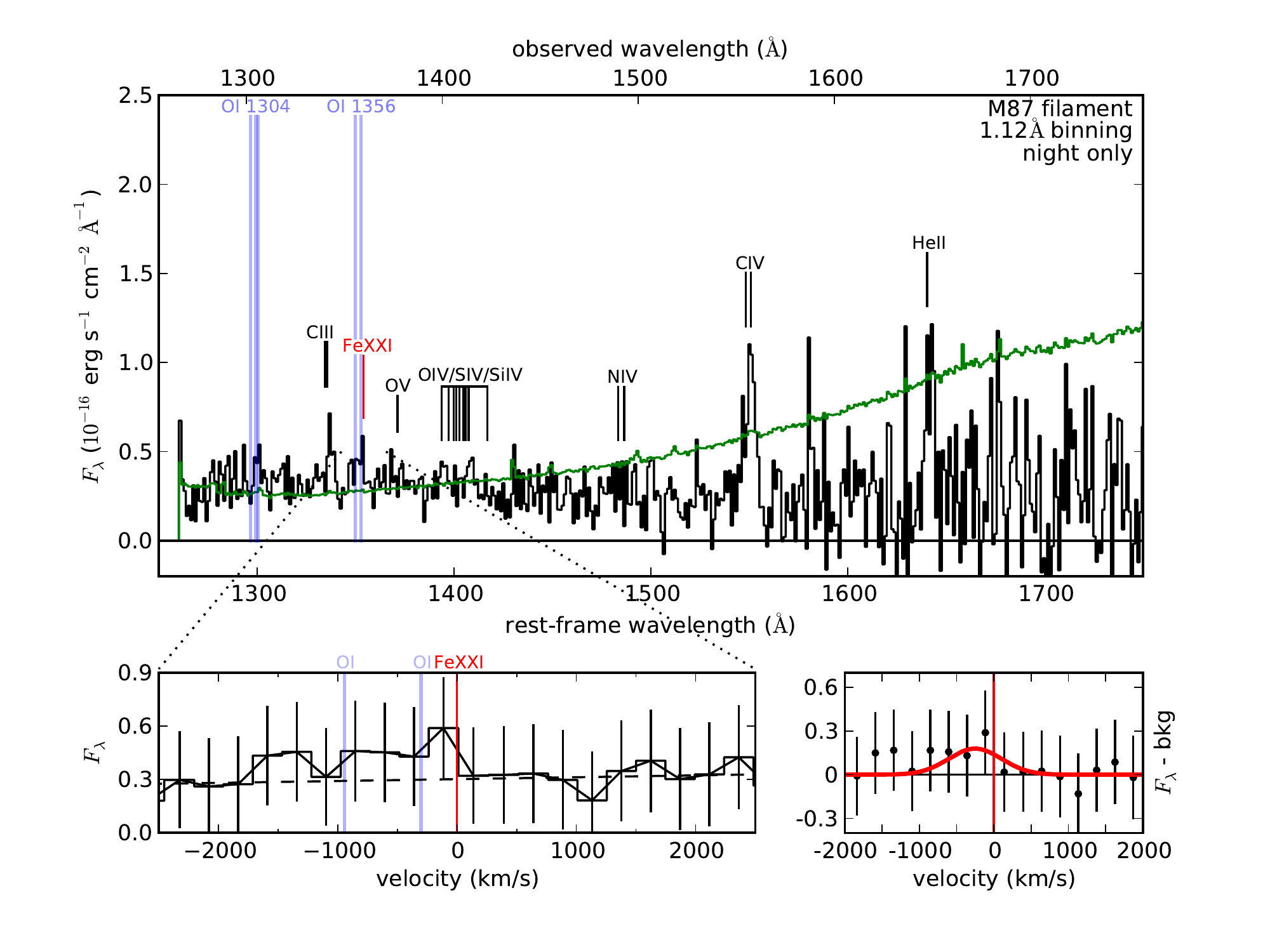}
\end{center}
\vspace{-0.8 cm}
\caption{(top) Average spectrum of COS G140L observations of a filament projected 1.9 kpc from the nucleus of M87. The spectrum is shown in black and the 1$\sigma$ error per bin is shown in green. The spectrum has been binned to a resolution of 14 pixels (approximately two resels). Prominent airglow lines are indicated in light blue. Since \ion{O}{i} $\lambda$1356 falls very close to the expected location of [\ion{Fe}{xxi}], we have filtered this observation to display only the spectrum taken during orbital night, which should eliminate any detectable emission from this airglow line (see Appendix). There is a weak suggestion of [\ion{Fe}{xxi}] in this spectrum, but it is not formally significant due to the large errors per bin. The lower left panel shows the spectrum within 2500 km/s of [\ion{Fe}{xxi}], as well as our linear fit to the continuum.  The lower right panel shows the background-subtracted spectrum within 2000 km/s of [\ion{Fe}{xxi}] as well as the best-fit Gaussian profile, which corresponds to a $1.0\sigma$ detection. The 90\% upper limit on the integrated [\ion{Fe}{xxi}] $\lambda$1354.1 flux is $1.7\times10^{-16}$ erg s$^{-1}$ cm$^{-2}$.}
\end{figure*}

Low-resolution G140L observations of one of the FUV-bright filaments are available, which we examine here. We downloaded and examined the \verb"x1dsum" files for datasets LBK104010, LC7T01010, and LC7T01010, which contain the merged spectrum as reduced by the calcos pipeline. At the redshift of M87, the geocoronal \ion{O}{i} $\lambda$1356 falls very close to the expected location of  [\ion{Fe}{xxi}] $\lambda$1354.1, so we first construct a spectrum from the data taken only from orbital night (using 7858 s out of a total exposure time of 31798 s) in order to exclude the possibility of geocoronal emission contaminating our search for [\ion{Fe}{xxi}]. We discuss this process in more detail in the Appendix.

We generate a series of bins approximately 14 pixels in size, and within each bin we average together the fluxes from each pixel of each spectrum that falls within the bin. A portion of the night-only spectrum from segment A is shown in Figure 2. The FWHM of the line spread function is about 7 pixels, so this binning combines about two resolution elements (resels) per bin. We use the \verb"calcos" pipeline errors for the error spectrum. We use a redshift of $v=1284\pm5$ km/s \citep{Cappellari2011} to estimate rest-frame wavelengths.

The \ion{He}{ii} multiplet and the \ion{C}{iv} doublet, which trace $10^5$ K gas, are clearly visible and have already been discussed by \citet{Sparks2012}. A weak positive residual is also visible at the expected location of [\ion{Fe}{xxi}]. We perform Gaussian fits to the observed [\ion{Fe}{xxi}] line profile, after subtracting a linear continuum fit to regions of the spectrum outside the line (Figure 2, lower left panel). We estimate likelihoods over a grid of three parameters (mean velocity $v_0$, velocity dispersion $\sigma$, and integrated flux $F$) using the $\chi^2$ statistic, and compute medians and central 68\% confidence intervals for each parameter, marginalizing over the other two. The best-fit profile has $v_0 = -250$ km/s, $\sigma = 356$ km/s, and $F= 7.2\times10^{-17}$ erg s$^{-1}$ cm$^{-2}$. The best-fit profile is shown in Figure 2 (lower right panel). However, the statistical significance of this detection is only 1.0$\sigma$, so it is not formally significant. The 90\% upper limit on the integrated flux is $1.7\times10^{-16}$ erg s$^{-1}$ cm$^{-2}$. Assuming a distance to M87 of 16.7 Mpc \citep{Blakeslee2009}, this corresponds to an [\ion{Fe}{xxi}] $\lambda$1354.1 line luminosity of $L < 5.8\times10^{36}$ erg s$^{-1}$ within our pencil-beam aperture (or if the emission is uniformly filling the field, this limit increases by about 15\%; see section 1.1).

We next tested the strength of the [\ion{Fe}{xxi}] feature as a function of the solar altitude relative to the instrument. We found that, with the exception of three observations in dataset LC7T02010 (which are discussed in the Appendix), the line flux remains constant, implying that \ion{O}{i} $\lambda$1356 airglow emission is negligible in these spectra (this is also true for the M87 nuclear spectra discussed in the next section). We therefore also consider the combination of all the M87 filament data (daytime and night-time, excepting the three contaminated observations mentioned above). This spectrum is shown in Figure 3. For this spectrum, we measure for  [\ion{Fe}{xxi}] best-fit line parameters of $v_0 = -390^{+230}_{-140}$ km/s, $\sigma = 400^{+70}_{-120}$ km/s, and $F= 7.8^{+3.8}_{-3.6}\times10^{-17}$ erg s$^{-1}$ cm$^{-2}$. These parameters are nearly identical to the night-only values. For \ion{O}{i} $\lambda$1356 with an expected 3:1 ratio between the lines in the doublet, the velocity relative to the position of [\ion{Fe}{xxi}] is about $-820$ km/s, so this line is about halfway between the expected locations for [\ion{Fe}{xxi}] and airglow emission. However, the night-only line also has a similar negative velocity, and there are additional uncertainties such as a 150 km/s uncertainty in absolute wavelength calibration, so we believe the line can properly be attributed to [\ion{Fe}{xxi}]. The negative velocity is a bit of a concern, however, and probably unphysical; follow-up observations at better spectral resolution (see section 8) will hopefully help clarify the situation. As mentioned above, the flux is nearly identical to the night-only case, but the longer integration time yields a larger S/N, so that the significance of the feature increases to 2.2$\sigma$ in this spectrum. 

\begin{figure*}
\begin{center}
\includegraphics[width=17cm]{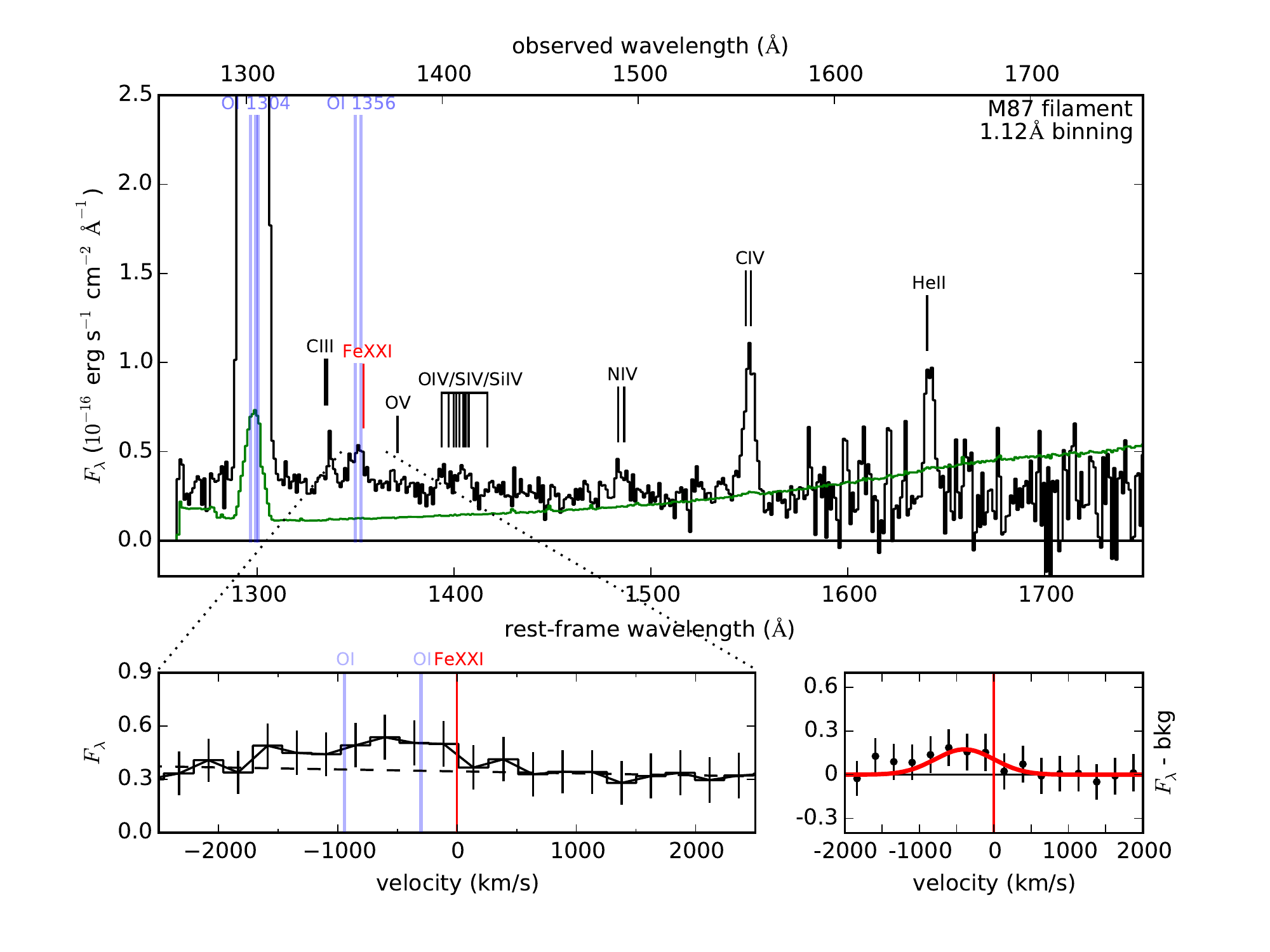}
\end{center}
\vspace{-0.8 cm}
\caption{(top) Average spectrum of COS G140L observations of a filament projected 1.9 kpc from the nucleus of M87. Unlike the previous Figure, we have not restricted this spectrum to the night-only observations; \ion{O}{i} $\lambda$1304 airglow is therefore extremely prominent, but we argue that the feature near \ion{O}{i} $\lambda$1356 is probably [\ion{Fe}{xxi}], with a negligible airglow contribution. The feature is significant at 2.2$\sigma$.}
\end{figure*}

This spectrum also contains information about [\ion{Fe}{xix}] $\lambda$1118.1. In Figure 4 we display the merged G140L spectrum of segment B, which covers the shorter-wavelength portion of the FUV, using all the observations from all three datasets targeting the filament region, as in Figure 3. Below 1100\AA{} the sensitivity of COS declines dramatically, and at 1134\AA{} \ion{N}{i} airglow emission is visible, but between these two regions the [\ion{Fe}{xix}] $\lambda$1118.1 line can be constrained. There is a weak feature at this location, significant at just $0.3\sigma$, with the best-fitting Gaussian having $v_0 = -190$ km/s, $\sigma = 346$ km/s, and $F = 1.6\times10^{-17}$ erg s$^{-1}$ cm$^{-2}$. The 90\% upper limit on the integrated flux is $8.6\times10^{-17}$ erg s$^{-1}$ cm$^{-2}$, corresponding to a line luminosity $L < 2.9\times10^{36}$ erg s$^{-1}$.

\begin{figure*}
\begin{center}
\includegraphics[width=17cm]{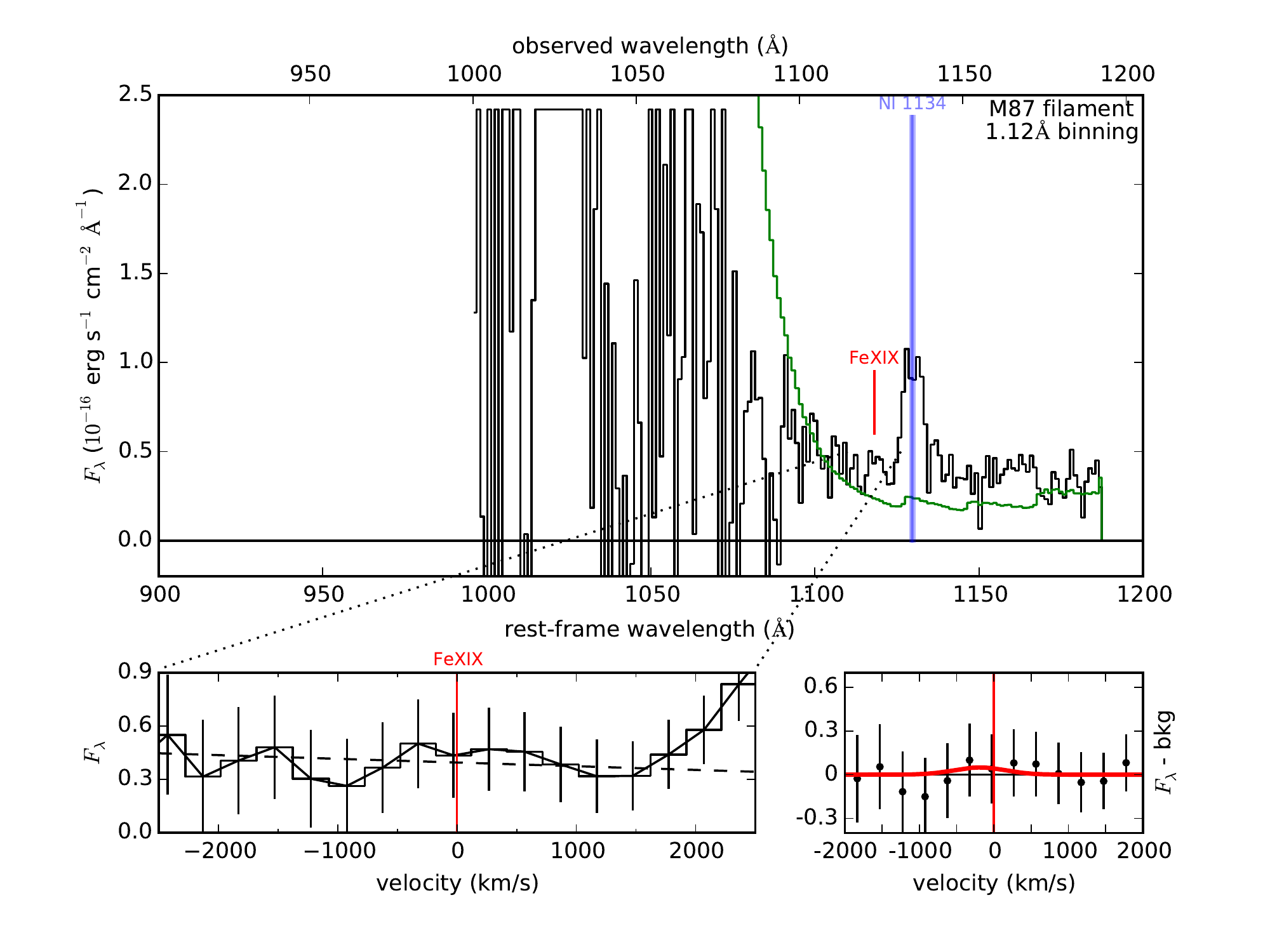}
\end{center}
\vspace{-1.1 cm}
\caption{(top) Average spectrum of the shorter-wavelength segment B for three COS G140L observations of the a filament projected 1.9 kpc from the nucleus of M87. The spectrum is shown in black and the 1$\sigma$ error per bin is shown in green. The spectrum has been binned to a resolution of 14 pixels (approximately two resels). Prominent airglow lines are indicated in light blue. The sensitivity of COS declines significantly below 1100\AA, so the spectrum is entirely noise-dominated below this wavelength. There is a small positive residual at the location of  [\ion{Fe}{xix}] in this spectrum, which is only significant at $0.3\sigma$. The lower left panel shows the spectrum within 2500 km/s of [\ion{Fe}{xix}], as well as our linear fit to the continuum.  The lower right panel shows the background-subtracted spectrum within 2000 km/s of [\ion{Fe}{xix}] as well as the best-fit Gaussian profile, which corresponds to a $0.3\sigma$ detection. The 90\% upper limit on the integrated [\ion{Fe}{xix}] $\lambda$1118.1 flux is  $0.9\times10^{-16}$ erg s$^{-1}$ cm$^{-2}$.}
\end{figure*}

In Figure 5, we illustrate an emission measure analysis based on these constraints.  Using the emissivities shown in Figure 1, we derive the pencil-beam emission measure $\int{n_e n_p dl} = L/(\epsilon \times A \times Z_{\text{Fe}})$ corresponding to the constraints on the line luminosities, where $A$ is the physical area subtended by our aperture at the distance of M87 ($3.0\times10^4$ pc$^2$) and $Z_{\text{Fe}}$ is the fractional abundance of Iron relative to Hydrogen in the hot plasma, which we assume to be $3.16\times10^{-5}$ (i.e. Solar abundance; \citealt{Grevesse1998}). We consider the 90\% upper limit for [\ion{Fe}{xix}] and the 2.2$\sigma$ measurement for [\ion{Fe}{xxi}], although we also show the 90\% upper limit on [\ion{Fe}{xxi}] from the night-only spectrum in order to be conservative. 

The resulting curves for [\ion{Fe}{xxi}] have a clear minimum around $T = 10^{7.05}$ K, which is where the emissivity of [\ion{Fe}{xxi}] is highest. This is the portion of the plot at which one would naturally expect physical solutions.The implied emission measure at this temperature is $1.4\pm0.6 \times10^{20}$ cm$^{-5}$.  [\ion{Fe}{xix}] has a lower emissivity than [\ion{Fe}{xxi}], so it does not give as strong of a constraint; the 90\% upper limit on the emission measure at its peak temperature is $5.9\times10^{20}$ cm$^{-5}$.

Next we estimate the physical properties of the hot gas in our pencil-beam aperture. To do this, we combine our constraints with an additional piece of information: the pressure of the gas as measured from X-ray observations. We adopt a value for the pressure of $P = 1.91 P_e = 4\times10^6$ cm$^{-3}$ K, based on \citet{Churazov2008}. This value is fairly robust: \citet{DiMatteo2003}, \citet{Churazov2008}, and \citet{Russell2015} all find similar results, and they all find that the deprojected pressure profile is approximately constant within the central 2.5 kpc, except for an increase in the central 100 pc.

Combining our results for the emission measure with the X-ray measurements of the pressure, we can derive several additional constraints. The equivalent electron density is $n_e = P_e/T$, which we plot in the upper axis of Figure 5. The corresponding electron column in our aperture is given by 

\begin{align}
\begin{split}
\label{eq_Ne}
N_e &= n_e l =  1.6\times10^{21} \text{ cm}^{-2}  \times \left(\frac{\text{EM}}{2.5\times10^{20} \text{ cm}^{-5} }\right) \\
 &\times \left(\frac{T}{10^{7.05} \text{ K}}\right) \times \left(\frac{4 \times10^6 \text{ cm}^{-3} \text{ K}}{P}\right)\times \left(\frac{Z_{\odot}}{Z}\right)
\end{split}
\end{align}

\noindent where the coefficient is derived by assuming $p = 1.91 p_e$ and a proton-to-electron ratio of 0.83. Note that this implies that [\ion{Fe}{xxi}] and [\ion{Fe}{xix}] would indeed be optically thin, as we had assumed when deriving the emissivity. The corresponding effective path length of [\ion{Fe}{xxi}]-emitting gas in our aperture is 

\begin{align}
\begin{split}
\label{eq_l}
l &=  3.0 \text{ kpc}  \times \left(\frac{\text{EM}}{2.5\times10^{20} \text{ cm}^{-5} }\right) \\
 &\times \left(\frac{T}{10^{7.05} \text{ K}}\right)^2 \times \left(\frac{4 \times10^6 \text{ cm}^{-3} \text{ K}}{P}\right)^2\times \left(\frac{Z_{\odot}}{Z}\right)
 \end{split}
\end{align}

\noindent We use these relations to plot curves indicating $N_e$ and $l$ in Figure 5. The total mass in hot gas traced by [\ion{Fe}{xxi}] is $M = \mu_e m_p N_e A$, where $A$ is the physical area subtended by our aperture at the distance of M87 and we assume $\mu_e = 1.2$. The metallicity dependence enters through the emission measure, which scales as $(Z/Z_{\odot})^{-1}$; if we assume the pressure is robustly determined, then the path length, column density, and hot gas mass all scale as $(Z/Z_{\odot})^{-1}$, as shown in equations \ref{eq_Ne} and \ref{eq_l}.

\begin{figure}
\begin{center}
\includegraphics[width=8.5cm]{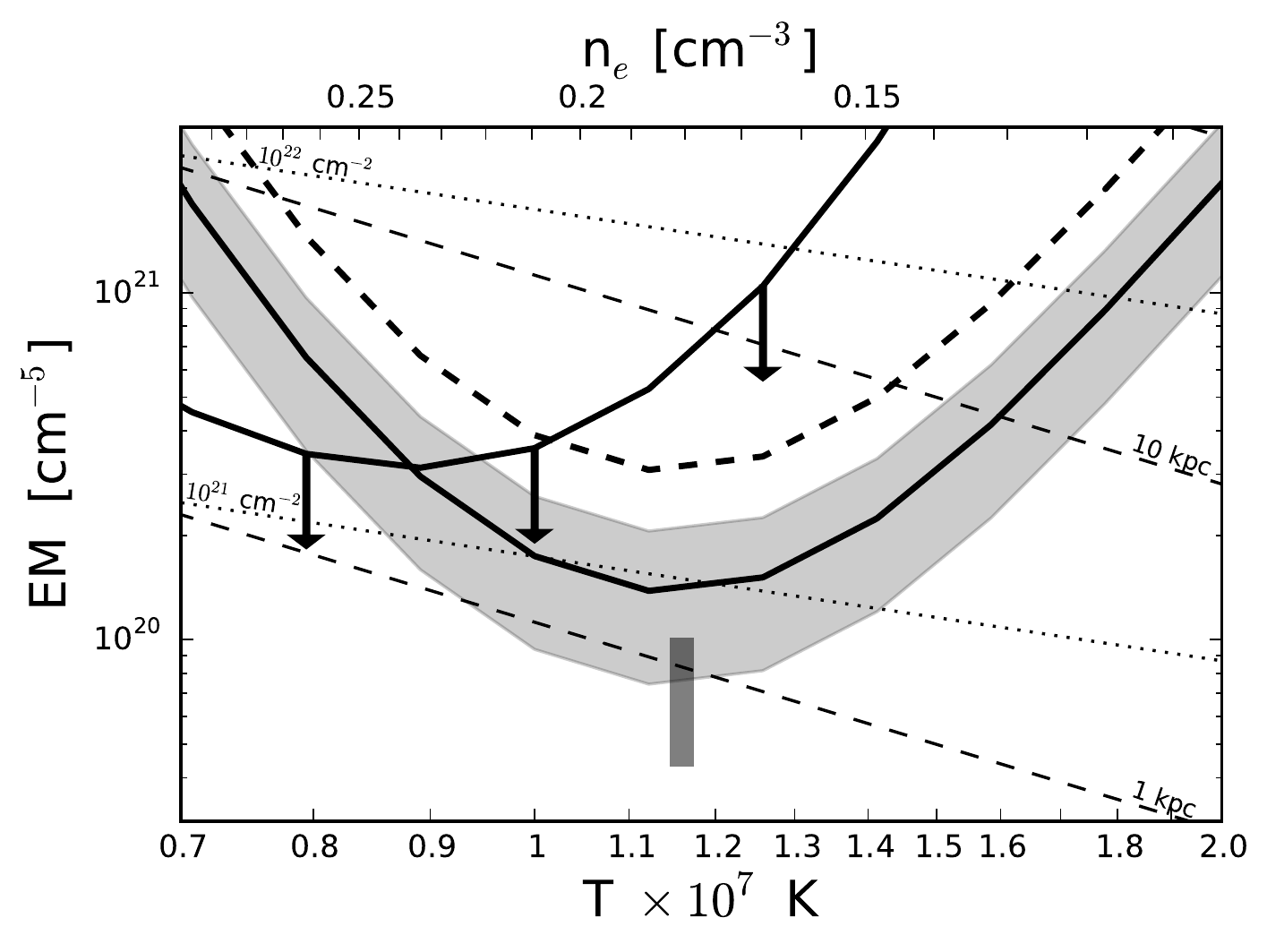}
\end{center}
\vspace{-0.4 cm}
\caption{Constraints on the pencil-beam emission measure $\int{n_e n_p dl}$ from the filament sightline towards M87 inferred from our flux measurements for [\ion{Fe}{xxi}] (solid curve and shaded region for the measured value and 1$\sigma$ uncertainties for the spectrum in Figure 3; the dashed curve shows the 90\% upper limit from the night-only spectrum for comparison) and [\ion{Fe}{xix}] (upper left curve). We also compute the corresponding average electron density (upper X-axis), electron column density (dotted lines), and path length (dashed lines) for the hot gas, assuming a constant total pressure of $P = 1.91 P_e = 4\times10^6$ cm$^{-3}$ K in the sightline \citep{Churazov2008}. [\ion{Fe}{xix}] has a lower emissivity than [\ion{Fe}{xxi}], so it does not yield as strong of a constraint. The [\ion{Fe}{xxi}] curve is sharply peaked at $10^{7.05}$ K, and if the [\ion{Fe}{xxi}]-emitting gas has this temperature, we obtain physically plausible parameters for the hot gas in the sightline. The shaded dark rectangular region indicates the measured emission measure of 1 keV gas from the X-ray observations of \citet{Werner2013}, which we have scaled to an assumed Solar metallicity. This measurement is consistent with our estimate of the [\ion{Fe}{xxi}] luminosity. The physical values in this plot are all computed assuming Solar abundance, and scale with metallicity as discussed in the text.}
\end{figure}

At $10^{7.05}$ K, our estimates for these quantities are $N_e = 9\pm4 \times 10^{20}$ cm$^{-2}$, $l = 1.5\pm0.7$ kpc, and $M = 2.8\pm1.2 \times10^5 M_{\odot}$. These values are physically plausible for the gas around $10^7$ K in M87, and they also agree with X-ray measurements. For example, \citet{DiMatteo2003}, \citet{Churazov2008}, and \citet{Russell2015} all find deprojected electron density profiles which are fairly constant from 100 pc out to a few kpc, at values of 0.1-0.2 cm$^{-3}$. Additionally, the sightline is projected 1.9 kpc from the center of M87, so an effective path length of this order is a reasonable first-order assumption. 

There is also an X-ray measurement of the emission measure of 1 keV gas, which is an independent measurement against which we can compare our result. This measurement comes from the 1 keV component of the three-temperature spectral modeling for this region performed by \citet{Werner2013}. From their high-resolution Chandra map, they estimate the emission measure of 1 keV gas from the sightline corresponding to our observation to be around $2-4\times10^{60}$ cm$^{-3}$ arcsec$^{-2}$, using a metallicity of $1.46Z_{\odot}$ relative to the abundance tables of \citet{Grevesse1998}. This corresponds to an  [\ion{Fe}{xxi}] $\lambda$1354.1 flux of $3-5\times10^{-17} (Z/Z_{\odot})$ erg s$^{-1}$ cm$^{-2}$.  This is within 1$\sigma$ of our estimate of $7.8\pm3.6\times10^{-17}$ erg s$^{-1}$ cm$^{-2}$, assuming Solar abundance.

\subsection{\ion{C}{iv} and \ion{He}{ii}}

It is interesting to compare the lines tracing $10^7$ K gas to the stronger \ion{C}{iv} and \ion{He}{ii} lines which are characteristic of $10^5$ K gas in this spectrum. We fit a Gaussian to \ion{He}{ii} $\lambda$1640.4 and a double Gaussian (with normalizations fixed to the expected 2:1 ratio) to \ion{C}{iv} $\lambda$1548.2 and \ion{C}{iv} $\lambda$1550.8 in the full G140L filament spectrum (i.e. without the filtering to exclude data taken during orbital day). We measure velocity dispersions of around 600 km/s for both lines, which are far larger than the expected thermal velocities for C and He ions in $10^5$ K gas (10 and 20 km/s respectively). We interpret these dispersions as effects of spatial broadening, as discussed in the next section. 

Both lines also appear redshifted relative to the velocity of M87, by about 150 km/s. The absolute wavelength calibration of G140L is 150 km/s, so this apparent redshift could be explained by a wavelength miscalibration, but we think it is likely to be real. This filament has also been observed in H$\alpha$ \citep{Sparks1993} and [\ion{C}{ii}] $\lambda$158$\mu$m \citep{Werner2013}, and both observations show the same redshift for the filament. This bulk velocity corresponds to $\sim$40\% of the sound speed of the hot $10^7$ K ambient medium, which is evidence that bulk motions through the hot medium can be considerable. Interestingly, the weak [\ion{Fe}{xxi}] feature and the weaker [\ion{Fe}{xix}] feature both show no sign of redshift relative to M87; if these lines are confirmed, this would imply the $10^7$ K gas is not moving relative to the galaxy (i.e. the filaments are flowing through the hot medium). Deeper observations with higher spectral resolution are necessary to confirm this, however.

Finally, we consider the fluxes of these lines. We measure integrated fluxes of  $3.6\pm0.9 \times10^{-16}$ erg cm$^{-2}$ s$^{-1}$ and $3.6 \pm2.0\times10^{-16}$ erg s$^{-1}$ cm$^{-2}$ respectively for \ion{C}{iv} and \ion{He}{ii}, which are roughly consistent with the measurements of \citet{Sparks2012}. In Figure 6 we show the CIE emissivities for these two lines. They both peak at around $10^5$ K, but for Solar abundances their expected ratio at this temperature is about 110, much higher than the observed ratio of 1.0. This places a constraint on the shape of the differential emission measure (DEM) in this region: it must have far more weight towards gas at temperatures of $2\times10^6$ K and above, as compared to $10^5$ K.

\begin{figure}
\begin{center}
\includegraphics[width=8.5cm]{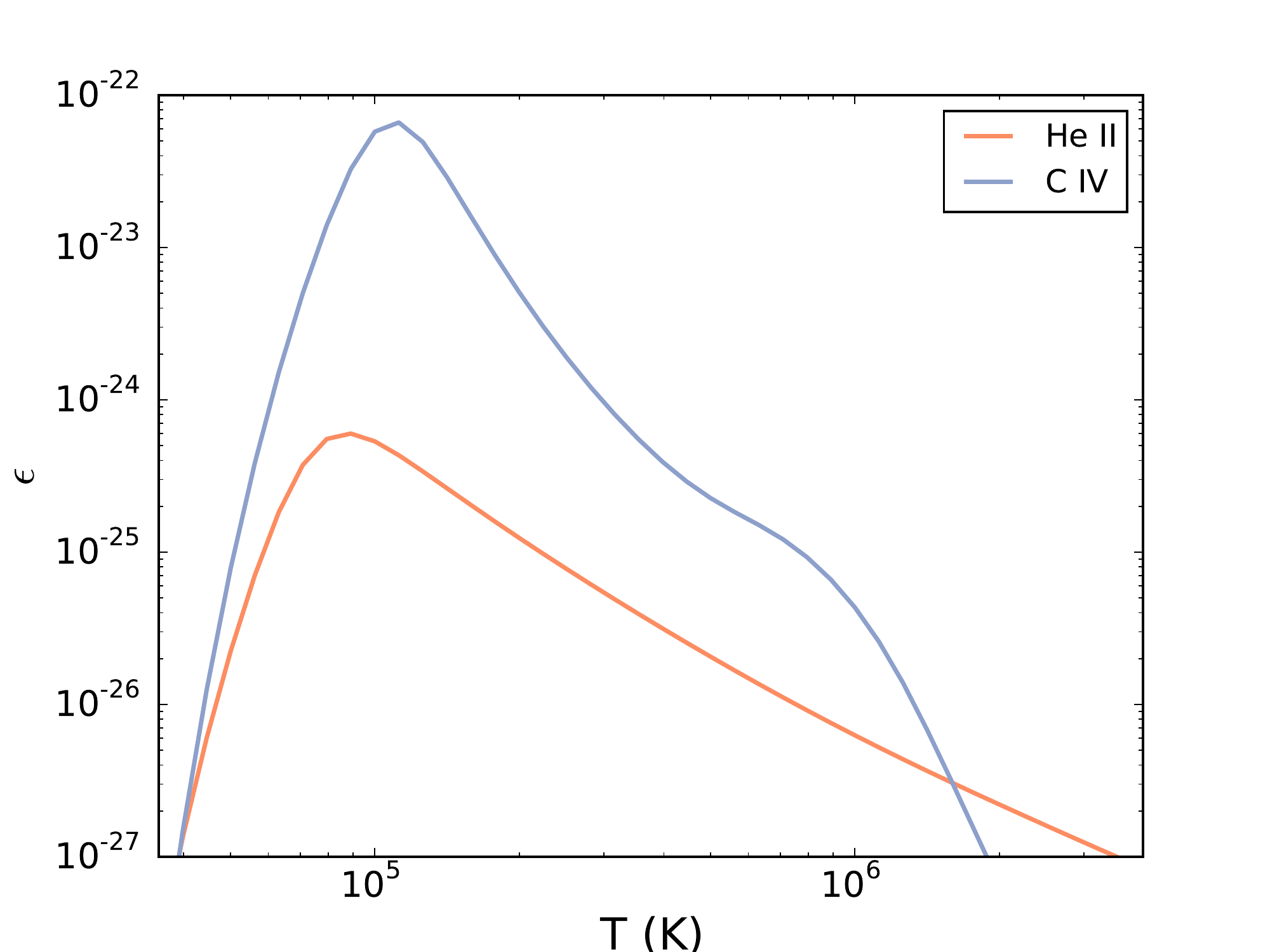}
\end{center}
\vspace{-0.3 cm}
\caption{Estimated emissivities (erg cm$^3$ s$^{-1}$) for the sum of the emission from the \ion{C}{iv} $\lambda1549$ doublet (blue) and the sum of the \ion{He}{ii} $\lambda$1640 multiplet (red), assuming Solar abundance and CIE. The predicted ratio of these lines at $10^5$ K is 110; to get the observed ratio of $\sim1.0$ a significant contribution from hotter gas is required. The ratio from cooling flow models is still too large by nearly an order of magnitude, but thermal conduction models seem to be able to reproduce the observed ratio \citep{Sparks2012}.  }
\end{figure}

Thermal conduction is a likely explanation for this unusual line ratio. A cooling flow is an alternative hypothesis, but a cooling should have a DEM which is inversely proportional to the cooling function \citep{Fabian1994}, which could only reduce the \ion{C}{iv} / \ion{He}{ii} ratio by an order of magnitude; this is insufficient to match the observed ratio. The thermal conduction model explored by \citet{Sparks2009} also significantly underpredicts \ion{He}{ii}, although their updated model has more success \citep{Sparks2012}. Additionally, collisions with the hot electrons from the ambient medium could suppress the fraction of C$^{3+}$ \citep{Hicks2001}, while simultaneously boosting \ion{He}{ii} since the latter is composed of strong recombination lines (it is the H$\alpha$ transition for Hydrogen-like Helium). Whether or not the latter effect is necessary, the physical picture we infer is that \ion{C}{iv} traces the $10^5$ K gas, while \ion{He}{ii} has contributions from a wide range of temperatures. 

It is worth mentioning the possibility of photoionization being important as well, especially since photoionization commonly produces \ion{C}{iv} / \ion{He}{ii} ratios near unity \citep{Allen1998}. While the radiation field from stars and the hot gas in M87 is insufficient to affect the ionization balance significantly here, perhaps a burst of activity from the AGN in the recent past might have contributed the necessary photons. This underscores the difficulty in interpreting the filamentary line emission. Clearly the exact physical mechanisms producing the \ion{C}{iv} and \ion{He}{ii} are not yet fully understood, complicating a kinematic analysis of this region.

\section{M87 Nucleus}

A sightline with the medium-resolution G130M grating targets the central region of M87, including the active nucleus at the edge of the aperture but excluding the jet and the bright knot HST-1.  There are four G130M observations of this sightline, with central wavelengths of 1318\AA, 1309\AA, 1300\AA, and 1291\AA{} respectively, yielding a total integration time of 4401 s. We generate a series of bins approximately 21 pixels in size (i.e. three resels, or 0.21\AA), and within each bin we average together the fluxes from each pixel of each spectrum that falls within the bin. The merged G130M nuclear spectrum is shown in Figure 7.

This spectrum contains many well-known lines typical for photoionized or collisionally excited plasmas with $T_e \sim 10^4 - 10^5$ K , many of which have optical counterparts that have previously been studied with HST (e.g. \citealt{Ford1994}, \citealt{Harms1994}, \citealt{Macchetto1997}, \citealt{Walsh2013}). Of particular note, the emission lines (particularly \ion{N}{V} $\lambda$1240 doublet and the \ion{C}{ii} $\lambda$1336 triplet) are extremely broad and \ion{C}{ii} appears to have a blueshifted absorption component as well. Our interest in this paper is primarily with the [\ion{Fe}{xxi}] line, which uniquely traces the $10^7$ K gas (note that [\ion{Fe}{xix}] falls outside the spectral coverage of these G130M observations).

Using the merged spectrum from Figure 7, we repeat the Gaussian fitting procedure from Section 2 at the location of [\ion{Fe}{xxi}].  We find $v_0 = -10$ km/s, $\sigma = 290$ km/s, and $F = 1.4 \times10^{-16}$ erg s$^{-1}$ cm$^{-2}$. The best-fit profile is shown in Figure 7 (lower right panel). This is not a significant detection of [\ion{Fe}{xxi}] (S/N = 0.7). The 90\% upper limit on the integrated flux is $4.15\times10^{-16}$ erg s$^{-1}$ cm$^{-2}$, corresponding to a line luminosity of $L < 1.4\times10^{37}$ erg s$^{-1}$ within this sightline.

\begin{figure*}
\begin{center}
\includegraphics[width=17cm]{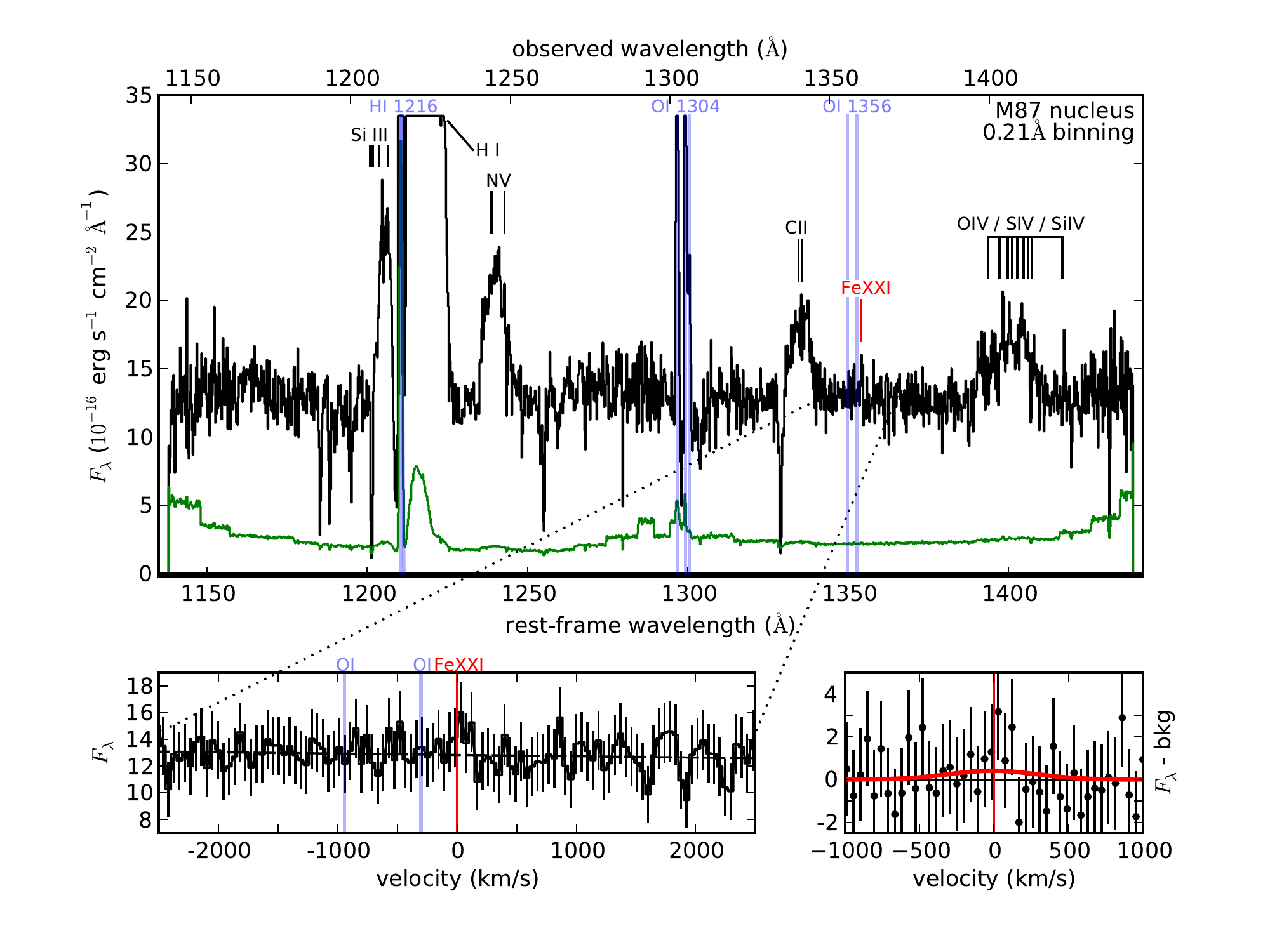}
\end{center}
\vspace{-1.2 cm}
\caption{ (top) Average spectrum of four COS G130M observations of the nucleus of M87. The spectrum is shown in black and the 1$\sigma$ error per bin is shown in green. The spectrum has been binned to a resolution of 21 pixels (approximately three resels). Prominent airglow lines are indicated in light blue. The Lyman-$\alpha$ aigrlow feature is blended with the extremely broad Lyman-$\alpha$ emission line from the nucleus of M87 and we truncate these lines for clarity. Other lines are also identified and marked; some of them are also extremely broad. The lower left panel shows the spectrum within 2500 km/s of [\ion{Fe}{xxi}], as well as our linear fit to the continuum. There is no evidence for geocoronal \ion{O}{i} contamination in this region. The lower right panel shows the background-subtracted spectrum within 1000 km/s of [\ion{Fe}{xxi}] as well as the best-fit Gaussian profile, which corresponds to a $0.7\sigma$ detection. The 90\% upper limit on the integrated [\ion{Fe}{xxi}] line flux is $4.15\times10^{-16}$ erg s$^{-1}$ cm$^{-2}$. }
\end{figure*}

In Figure 8, we repeat the emission measure analysis which we performed above for the filament, using the same pressure of $P = 1.91 P_e = 4\times10^6$ cm$^{-3}$ which we used for the filament region. The 1."25 radius of the COS aperture corresponds to 98 pc here, which is roughly the same size as the Bondi radius for M87 \citep{DiMatteo2003}, although the sightline is offset from the black hole so the plasma within the Bondi radius only covers a portion of our aperture. The pressure within the Bondi radius increases above our fiducial value, but the emission measure only rises logarithmically towards the center \citep{Russell2015}, so the contribution from the Bondi flow to [\ion{Fe}{xxi}] in this region is not likely to be dominant. 

\begin{figure}
\begin{center}
\includegraphics[width=8.5cm]{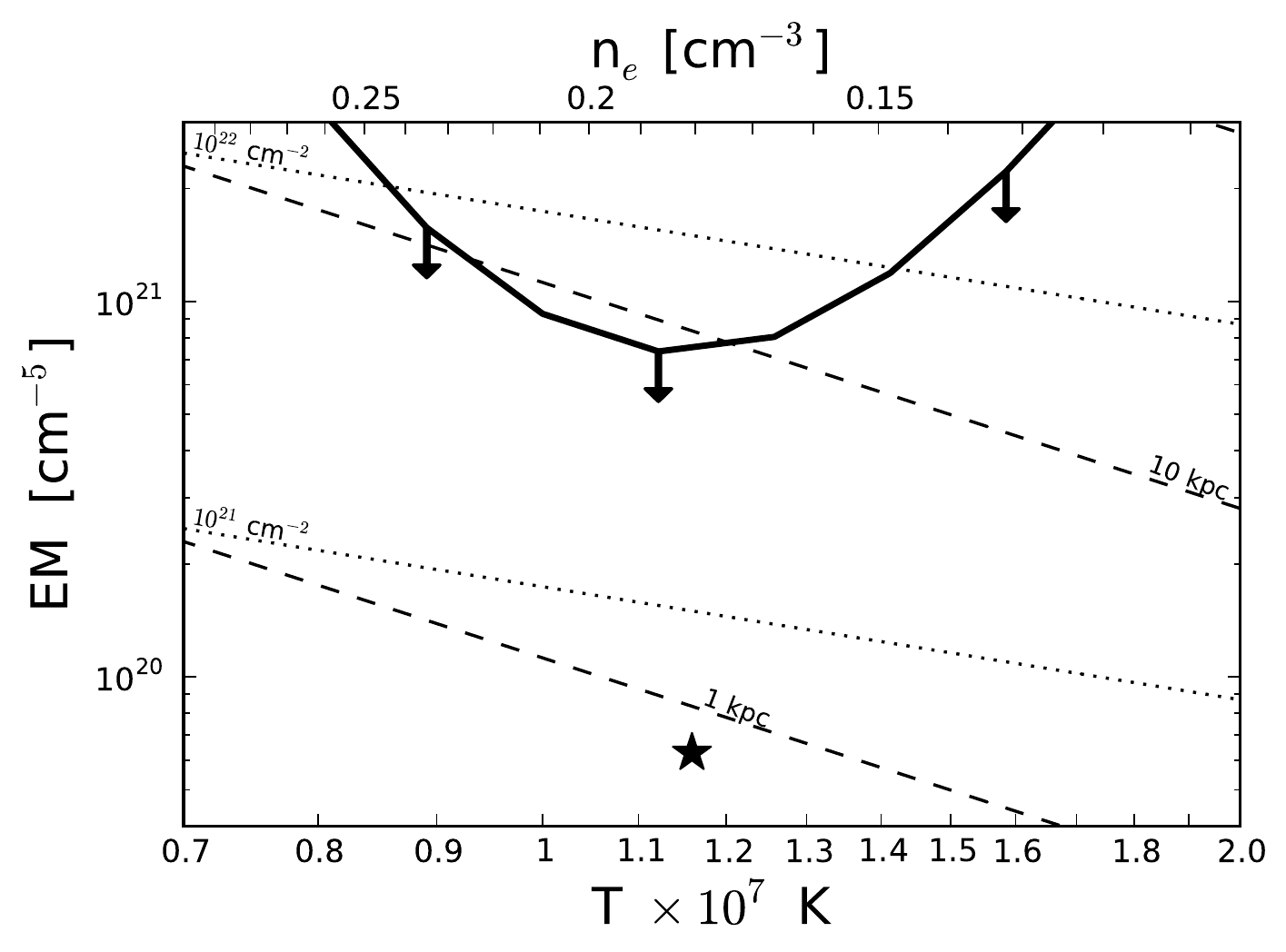}
\end{center}
\vspace{-0.3 cm}
\caption{The 90\% upper limit on the pencil-beam emission measure $\int{n_e n_p dl}$ from the nuclear sightline towards M87 inferred from our limits on the measured [\ion{Fe}{xxi}] flux. We also compute limits on the corresponding average electron density (upper X-axis), electron column density (dotted lines), and path length (dashed lines) for the hot gas, assuming a constant total pressure of $P = 1.91 P_e = 4\times10^6$ cm$^{-3}$ K in the sightline \citep{Churazov2008}. This curve is sharply peaked at $10^{7.05}$ K, and if the [\ion{Fe}{xxi}]-emitting gas has this temperature the limits approach physically plausible parameters for the hot gas in the sightline. The star indicates the approximate emission measure of 1 keV gas estimated from the X-ray observations of \citet{Werner2010}, which we have scaled to an assumed Solar metallicity. This is consistent with our limits. The physical values in this plot are all computed assuming Solar abundance, and scale with metallicity as discussed in the text.}
\end{figure}

We find that the 90\% upper limit on the minimum emission measure is $7.4 \times10^{20}$ cm$^{-5}$. The electron column is $N_e < 4.7 \times10^{21}$ cm$^{-2}$, the effective path length is $l < 8.2$ kpc, and the hot gas mass is $M < 1.5 \times10^7 M_{\odot}$ within our pencil-beam aperture.

There is also an X-ray measurement of the emission measure of 1 keV gas in this region, from the 5-temperature spectral modeling for this region performed by \citet{Werner2010}. From their high-resolution Chandra map, they estimate the emission measure of 1 keV gas from the nucleus of M87 to be around $6\times10^{60}$ cm$^{-3}$ arcsec$^{-2}$, assuming a distance of 16.1 Mpc and a metallicity of $1.3 Z_{\odot}$. Scaling this measurement to an assumed Solar metallicity, we predict an [\ion{Fe}{xxi}] $\lambda$1354.1 flux of $7\times10^{-17} (Z/Z_{\odot})$ erg s$^{-1}$ cm$^{-2}$ based on these X-ray observations. This is consistent with our observations.

\section{NGC 4696 Nucleus}

We also examine a sightline towards the center of NGC 4696. This is the brightest central galaxy of the Centaurus cluster, which is a classic cool-core system: the temperature of the intracluster medium decreases from a baseline of 3-4 keV down to a bit below 1 keV in the center (\citealt{Fukazawa1994}, \citealt{Allen1994}, \citealt{Allen2001}, \citealt{Molendi2002a}, \citealt{Fabian2005}, \citealt{Graham2006}, \citealt{Sanders2006}, \citealt{Sanders2008}, \citealt{Panagoulia2013}). As with M87, the X-ray spectra in the cool core prefer multi-temperature fits, and NGC 4696 also hosts extensive filamentary structure which radiates in optical emission lines (\citealt{Fabian1982}, \citealt{Crawford2005}). More recently,  [\ion{Fe}{x}] $\lambda$6374 emission has also been detected from this galaxy \citep{Canning2011}, which traces gas at $10^6$ K. \citet{Mittal2011} have also detected [\ion{C}{ii}] $\lambda$158$\mu$m, [\ion{O}{i}] $\lambda$63$\mu$m, and [\ion{N}{ii}] $\lambda$122$\mu$m emission which is largely coincident with the filamentary structure, suggesting the presence of $10^2$ K gas as well. The sightline towards this galaxy also covers its nucleus, which is an FR I radio source. 

The four datasets covering this sightline were taken in a similar configuration as the observation of the nucleus of M87, and have a total integration time of 5217 s. We average the four spectra and bin the resulting spectrum to 56 pixels (eight resels) per bin, in the same fashion as the nucleus of M87. NGC 4696 has a redshift of $z=0.009867$ ($v = 2958\pm15$ km/s; \citealt{DeVaucouleurs1991}), which we use to correct the spectra to the rest frame. Using the updated Planck cosmology \citep{Planck2015b} this gives a luminosity distance of 44.0 Mpc and implies a radius for the COS aperture corresponding to about 260 pc for this galaxy. 

The resulting merged spectrum is shown in Figure 9. This spectrum has a much lower continuum than the spectrum of the nucleus of M87; this is partially due to the higher distance of NGC 4696, and partially intrinsic. This observation is also much more heavily contaminated by airglow than the M87 observations: in addition to prominent \ion{H}{i} emission and strong emission from both \ion{O}{i} features, the \ion{N}{i} triplet at 1200\AA{} is also visible. Fortunately, the higher redshift of this galaxy means that none of these airglow lines fall near  [\ion{Fe}{xxi}]. 

We repeat the analysis we performed for M87. We fit a linear continuum to the region around the line.  Our best-fit Gaussian has $v_0 = 0$ km/s, $\sigma = 300$ km/s, and $F = 2.0\times10^{-17}$ erg s$^{-1}$ cm$^{-2}$, but it is only significant at $0.2\sigma$. The more relevant quantity is the 90\% upper limit on the integrated flux, which is $2.2\times10^{-16}$ erg s$^{-1}$ cm$^{-2}$ (corresponding to a line luminosity $L < 5.1\times10^{37}$ erg s$^{-1}$). We are somewhat sensitive to an accretion flow in this galaxy as well, but NGC 4696 has a less massive black hole than M87 and its Bondi radius fills a much smaller fraction of our aperture than in the M87 observation.

In Figure 10, we repeat the emission measure analysis which we performed above for M87. Here we assume a pressure of $2\times10^6$ cm$^{-3}$ K based on the deprojected density and temperature profiles at 1 kpc measured by \citet{Graham2006}. We could not find estimates in the literature against which to compare the emission measure within our aperture, but we find similar limits on the [\ion{Fe}{xxi}]-emitting column in NGC 4696 as compared to M87. The upper limit on the minimum emission measure is $3.1 \times10^{20}$ cm$^{-5}$. The electron column is $N_e < 4.0 \times10^{21}$ cm$^{-2}$, the effective path length is $l < 14.1$ kpc, and the hot gas mass is $M < 8.3 \times10^6 M_{\odot}$ within our pencil-beam aperture.

\begin{figure*}
\begin{center}
\includegraphics[width=17cm]{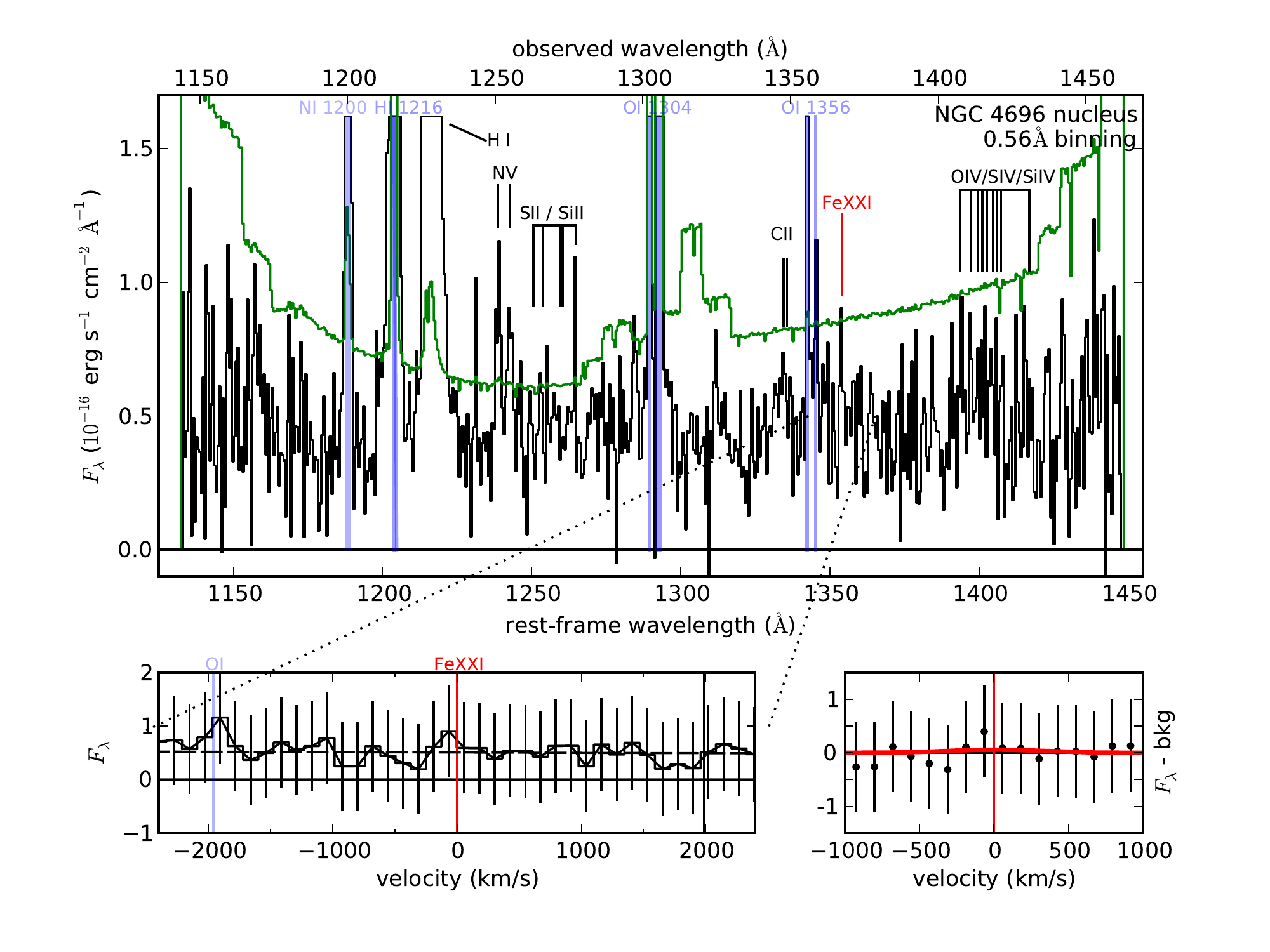}
\end{center}
\vspace{-1.0 cm}
\caption{ (top) Average spectrum of four COS G130M observations of the nucleus of NGC 4696. The spectrum is shown in black and the 1$\sigma$ error per bin is shown in green. The spectrum has been binned to a resolution of 56 pixels (approximately seven resels). Prominent airglow lines are indicated in light blue, and truncated for clarity; airglow contamination is much more significant in this spectrum than in the observations of M87. The lower left panel shows the spectrum within 2500 km/s of [\ion{Fe}{xxi}], as well as our linear fit to the continuum. There is weak geocoronal \ion{O}{i} emission in this region. The lower right panel shows the background-subtracted spectrum within 1000 km/s of [\ion{Fe}{xxi}] as well as the best-fit Gaussian profile, which corresponds to a $0.1\sigma$ detection. The 90\% upper limit on the integrated [\ion{Fe}{xxi}] line flux is $2.2\times10^{-16}$ erg s$^{-1}$ cm$^{-2}$. }
\vspace{0 cm}
\end{figure*}

\section{Potential Sources of Contamination}

In this section, we briefly explore other possible explanations for the positive residuals seen in our spectra towards M87 and NGC 4696. As discussed in Section 2 and in the Appendix, at the redshift of M87, [\ion{Fe}{xxi}] falls near the geocoronal \ion{O}{i} $\lambda1356$ lines, although in most of these observations the lines do not appear to contribute significantly.

It is also important to understand the continuum in these spectra, since the continuum emits much more flux than the forbidden lines we consider. M87 is one of the classic elliptical galaxies with an upturn in the FUV \citep{Bertola1980}. The exact cause of this "FUV upturn" is still unclear, but the likeliest possibility seems to be some type of hot evolved star, either isolated or in a binary system (see the review by \citealt{Oconnell1999} more discussion). Importantly, the FUV upturn, when it exists, appears not to have any prominent emission lines (\citealt{Ferguson1991}, \citealt{Brown1995}, \citealt{Brown1997}) and might even have a few broad absorption features \citep{Bica1996}.  We can therefore treat the continuum as being featureless, to first order, in the region around [\ion{Fe}{xxi}] and [\ion{Fe}{xix}]. NGC 4696 has a weaker continuum, but it likely has a similar origin.

Stellar coronae of some late-type stars also show emission in [\ion{Fe}{xxi}] $\lambda$1354.1, and in principle these may contribute to the flux in the lines of interest. In the Capella binary, each star has an [\ion{Fe}{xxi}] $\lambda$1354.1 luminosity of around $4\times10^{26}$ erg s$^{-1}$ \citep{Linsky1998}, so our upper limits correspond to $\sim 10^{10}-10^{11}$ Capella-type giant stars in each sightline, which is orders of magnitude above the expectation. We therefore do not expect stellar coronae in active binaries to contribute a major fraction of the [\ion{Fe}{xxi}] from these sightlines.

Observations should be able to separate the contributions from stars and gas. The stars in the nuclei of M87 have a velocity dispersion of 400-450 km/s \citep{Gebhardt2011} and in NGC 4696 they have a velocity dispersion of 250 km/s (HyperLEDA; \citealt{Makarov2014}) respectively. Our best-fit Gaussian profiles are broad, but in the M87 nucleus we do not see this level of broadening, and deeper observations should be able to distinguish a stellar velocity dispersion from the smaller velocities expected from Iron ions.

Finally, in nearby stellar coronal spectra that show the [\ion{Fe}{xxi}] line, the nearby \ion{C}{i} $\lambda$1354.3 line can often be an issue. Based on non-detections of the other \ion{C}{i} lines in these spectra it is unlikely that \ion{C}{i} $\lambda$1354.3 is significant here, and if it were present we would again expect it to have the velocity dispersion of the stars.

 On the other hand, for an early-type galaxy without an FUV upturn or any $10^7$ K gas, extragalactic observations which target the [\ion{Fe}{xxi}] emission from coronally-active late-type stars are conceivable. Such observations would be a unique constraint on the properties of these stars in elliptical galaxies.

\begin{figure}
\begin{center}
\includegraphics[width=8.5cm]{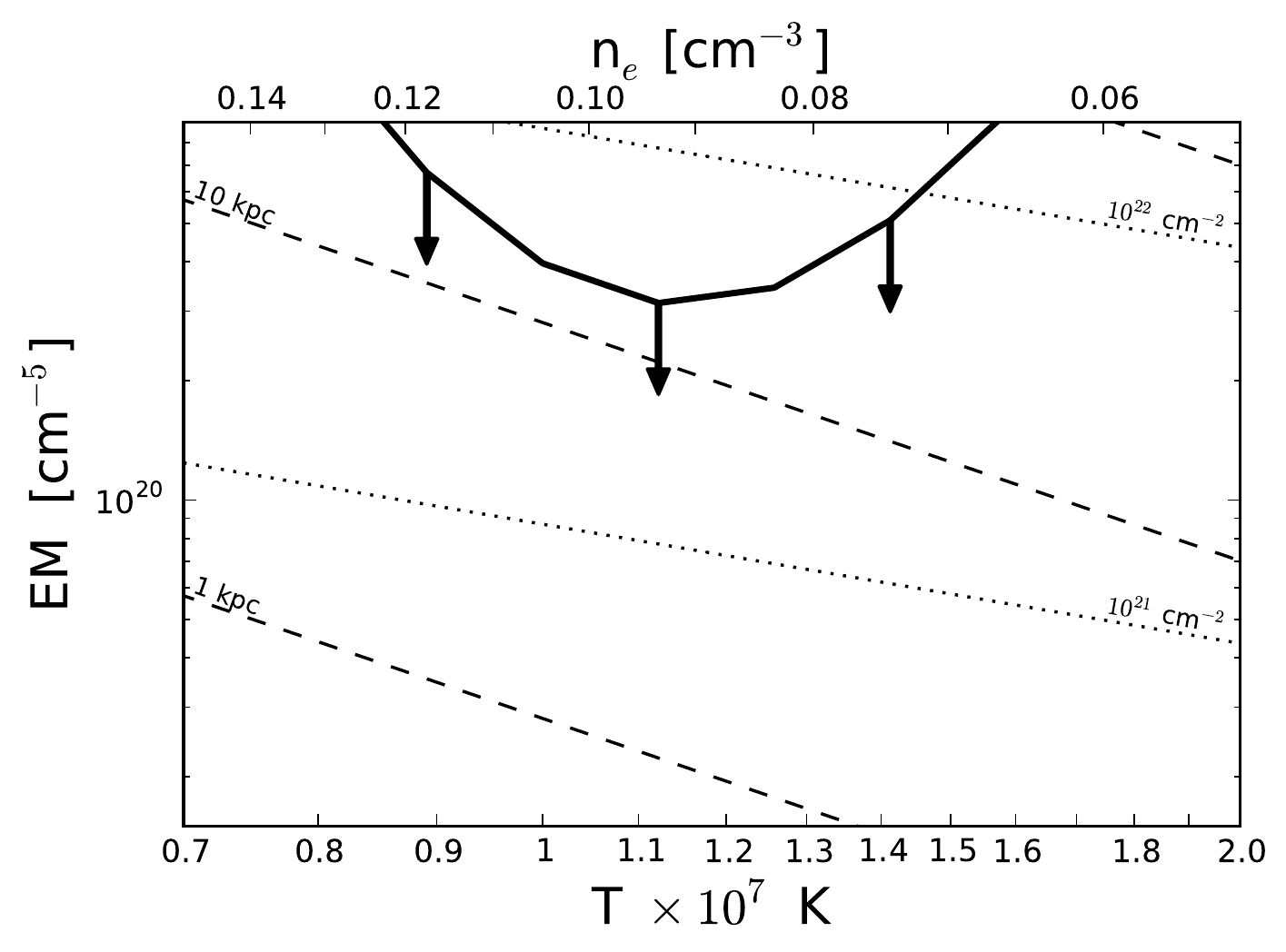}
\end{center}
\vspace{-0.3 cm}
\caption{The 90\% upper limit on the pencil-beam emission measure $\int{n_e n_p dl}$ from the nuclear sightline towards NGC 4696 inferred from our limits on the measured [\ion{Fe}{xxi}] flux. We also compute limits on the corresponding average electron density (upper X-axis), electron column density (dotted lines), and path length (dashed lines) for the hot gas, assuming a constant total pressure of $ P = 1.91 P_e = 2\times10^6$ cm$^{-3}$ K in the sightline \citep{Graham2006}. This curve is sharply peaked at $10^{7.05}$ K, and if the [\ion{Fe}{xxi}]-emitting gas has this temperature the limits approach physically plausible parameters for the hot gas in the sightline. The physical values in this plot are all computed assuming Solar abundance, and scale with metallicity as discussed in the text.}
\end{figure}

\section{Other Sightlines}

Here we briefly consider additional sightlines towards cooling flows or filaments seen in massive galaxy clusters. These clusters are all hotter than 1 keV, so they are not expected to host significant reservoirs of $10^7$ K gas, but our sightlines do probe regions of multiphase cooling so some $10^7$ K gas likely exists. We found publicly available COS observations towards A1795 ($z = 0.0619$; \citealt{McDonald2014}), Perseus ($z = 0.0176$, \citealt{Churazov2003}), Phoenix (SPT-CL J2344-4243; $z = 0.596$; \citealt{McDonald2012}), and Zw 3146 ($z = 0.2906$, \citealt{Allen1992}). We also consider the COS observations of a shock front in HCG 92 ($z = 0.021$; \citealt{White1999}), which is likely cooler than $10^7$ K, but may contain a hotter component at $10^7$ K. 

The sightline towards A1795 targets a filament projected about 30 kpc from the central galaxy. According to deep X-ray observations \citep{Ehlert2015}, the filament can be fit with a multi-temperature model, but the hotter component is around 4 keV and the cooler component is about 2 keV, which is not quite ideal for producing these FUV lines. However, the COS G140L spectrum shows a weak positive residual at [\ion{Fe}{xxi}], significant at 0.7$\sigma$, with a 90\% upper limit on the integrated flux  of $1.3\times10^{-16}$ erg s$^{-1}$ cm$^{-2}$. The deprojected electron pressure at the location of this sightline is around $8\times10^5$ cm$^{-3}$ K (\citealt{Tamura2001}, \citealt{Ettori2002}); with this assumption and an assumed Solar metallicity our upper limit on the effective path length of gas at $10^{7.05}$ K is $l <23$ kpc and the upper limit on the electron column density is $N_e < 5\times10^{21}$ cm$^{-2}$. 

In the A1795 spectrum there is also a weak detection of \ion{O}{vi} $\lambda$1038 claimed by \citet{McDonald2014}. They place a upper limit of 90 km/s on the width of the line, which would imply that the line is not significantly spatially broadened (recall from section 1.1 that aperture-filling emission observed with G140L can be expected to have $\sigma \sim 650$ km/s). At the redshift of A1795, the COS aperture subtends a diameter of about 3 kpc, so if the filament is significantly narrower than this, spatial broadening might not be important. On the other hand, the thermal velocity of O ions at $10^5$ K is only about 10 km/s, so there remains the possibility of up to 90 km/s of turbulent broadening and/or spatial broadening in this feature. 

In Perseus we examine a deep observation of the so-called outer ``horseshoe'' filament projected about 20 kpc from the center of NGC 1275. The temperature of the intracluster medium at this location is around 3 keV, much too hot to expect emission from [\ion{Fe}{xxi}], and indeed we see no trace of this line in the combined G140L spectrum. The 90\% upper limit on the flux is $6.2\times10^{-17}$ erg s$^{-1}$ cm$^{-2}$. Combining this with the estimated electron pressure of $1.4\times10^6$ cm$^{-3}$ K \citep{Churazov2003} at this location, we derive an upper limit on the effective path length of gas at $10^{7.05}$ K of $l < 3$ kpc and an upper limit on the electron column density of $N_e < 1\times10^{21}$ cm$^{-2}$, which are both fairly tight limits. 

The Phoenix cluster is located at much higher redshift than the other targets we have considered so far. This allows us to constrain some of the other lines in Figure 1, which lie at wavelengths too short for COS to observe with good sensitivity in the local Universe. As long as our source fills the COS aperture, the $1/R^2$ decrease in the flux is matched by a $1/R^2$ increase in area, so our sensitivity remains roughly constant (except for a small decrease stemming from the different redshift dependences of the angular diameter distance and the luminosity distance), allowing us to derive useful constraints even at moderate redshifts. 

The temperature of this supermassive cluster is around 13 keV \citep{McDonald2012}, but the COS observations target the powerful (${\dot M} \sim 3800 M_{\odot}$ yr$^{-1}$; \citealt{McDonald2012}) cooling flow. X-ray observations of the cooling flow show a multiphase medium, but they do not detect any gas below 1.8 keV, with upper limits on the cooling rate below this temperature on the order of hundreds of $M_{\odot}$ yr$^{-1}$ \citet{Tozzi2015}. The G160M spectrum covers [\ion{Fe}{xviii}] and [\ion{Fe}{xxiii}], but neither line is detected. Our 90\% upper limits on the flux are $1.7\times10^{-16}$ erg s$^{-1}$ cm$^{-2}$ and $1.9\times10^{-16}$ erg s$^{-1}$ cm$^{-2}$ respectively. These limits are comparable to the X-ray limits from the \citet{Tozzi2015} RGS spectrum. While a detailed comparison would be outside of the scope of this paper, the ratio of \ion{Fe}{xxi} $\lambda$12.16 to [\ion{Fe}{xxi}] $\lambda$1079 is about 50 at $10^7$ K, so our upper limit corresponds to an upper limit of $1\times10^{-14}$ erg s$^{-1}$ cm$^{-2}$ for the X-ray line. From Fig. 6 of \citet{Tozzi2015}, the integrated flux of this line is sightly under this limit (we estimate $4\times10^{-15}$ erg s$^{-1}$ cm$^{-2}$). The non-detection of \ion{Fe}{xxi} $\lambda$9.48, on the other hand, implies a [\ion{Fe}{xxi}] $\lambda$1354.1 limit of around $6\times10^{-15}$ erg s$^{-1}$ cm$^{-2}$. Our COS limits are much deeper than this, so if this redshifted line fell in the G160M bandpass COS would be able to improve considerably the constraint on $10^7$ K gas in the cool core of Phoenix. 

Zw 3146 also has a moderate redshift and its G140L spectrum has good sensitivity for [\ion{Fe}{xxi}], [\ion{Fe}{xix}], and [\ion{Fe}{xxiii}]. This cluster has a core temperature of around 3 keV \citep{Diehl2008} and a cooling flow nearly as strong as Phoenix \citep{Edge1994}, and the COS sightline passes through this cooling flow.  These three lines are not detected in this spectrum; the respective 90\% upper limits are $2.4\times10^{-16}$ erg s$^{-1}$ cm$^{-2}$, $1.2\times10^{-16}$ erg s$^{-1}$ cm$^{-2}$, and $0.8\times10^{-16}$ erg s$^{-1}$ cm$^{-2}$. Again we do not have a good constraint on the hot gas pressure within this sightline, so we simply report the upper limits on the flux here.

Finally, HCG 92 (Stefan's Quintet) has a diffuse intragroup medium with $kT \approx 0.5$ keV \citep{Trinchieri2003}, and a large X-ray bright shock near NGC 7318b which is the target of the COS observations. Temperature measurements are often difficult in shocked plasma, but \citet{Trinchieri2005} infer a multi-temperature medium with the hotter component around 1 keV while \citet{OSullivan2009} find a single-temperature model at about 0.6 keV is sufficient. We combine all the available COS data, which targets multiple sightlines along the X-ray shock front, and find no evidence for any of our high-excitation FUV lines. Unfortunately [\ion{Fe}{xxi}] lies in the gap between the G130M and the G160M observations, but we obtain 90\% upper limits for [\ion{Fe}{xix}] and [\ion{Fe}{xxiii}] of $3.7\times10^{-16}$ erg s$^{-1}$ cm$^{-2}$ and $7.4\times10^{-16}$ erg s$^{-1}$ cm$^{-2}$ respectively. Given the uncertain properties of the gas in the shock, the pressure is not well-constrained, and we simply report the upper limits on the flux. 

In Table 3 we collect many of the constraints reported in this work. For each FUV line we examine, we list the constraint on the flux and the associated pencil-beam emission measure.The latter quantity depends on the metallicity and temperature of the gas; we adopt the same convention as used above in our emission measure analyses and assume Solar abundance and that each species is at the temperature corresponding to its peak emissivity. Defined in this way, the quoted emission measure can be derived from the flux without any information other than an assumed distance to the target, which we derive from the quoted redshift in every case except for M87, where we use a distance of 16.7 Mpc \citep{Blakeslee2009}. In the systems where we have an estimate of the hot gas pressure, we also list the derived constraint on the effective path length of hot gas and the free electron column density.

\begin{table*}
\begin{minipage}{150mm}
\caption{90\% Upper Limits Reported in this Work}
\begin{tabular}{llcccc}
\hline
Target & Line & Integrated Flux & $\int{n_e n_p dl}$ & $l$ & $N_e$\\
 & & ($10^{-16}$ erg s$^{-1}$ cm$^{-2}$) & ($10^{20}$ cm$^{-5}$) & (kpc)& ($10^{21}$ cm$^{-2}$)\\
 \hline 
M87 filament (night-only)& [\ion{Fe}{xxi}] & 1.7 & 3.2 &3.5& 2\\
M87 filament (day+night)* & [\ion{Fe}{xxi}] & $0.8\pm0.4$ & $1.4\pm0.6$ &$1.5\pm0.7$& $9\pm4$\\
M87 filament (day+night) & [\ion{Fe}{xix}] & 1.5 & 5.9 &3.8 &3\\
M87 nucleus & [\ion{Fe}{xxi}] & 4.2 & 8.0 &8.2 &5\\
NGC 4696 nucleus & [\ion{Fe}{xxi}] & 2.2 &4.1 &14.1 &40\\
A1795 filament & [\ion{Fe}{xxi}] & 1.3 &2.9 &23& 50\\
Perseus filament & [\ion{Fe}{xxi}] & 0.6 &1.2&3 &10\\
Phoenix nucleus & [\ion{Fe}{xviii}] & 1.7 & 23&-& -\\
Phoenix nucleus & [\ion{Fe}{xxiii}] & 1.9 &310&- &- \\
Zw 3146 nucleus & [\ion{Fe}{xxi}] & 2.4 &12 &-&- \\
Zw 3146 nucleus & [\ion{Fe}{xix}] & 1.2 &12&- & -\\
Zw 3146 nucleus & [\ion{Fe}{xxiii}] & 0.8 &55& -& -\\
HCG 92 shock & [\ion{Fe}{xix}] & 3.7 & 13 & - & - \\
HCG 92 shock & [\ion{Fe}{xxiii}] & 7.4 & 27 & - & - \\
\hline
\end{tabular}
\\
\small{90\% upper limits reported in this work. For each line we constrain along each sightline, we present the 90\% upper limit on the integrated line flux as well as the upper limit on the pencil-beam emission measure. For the latter quantity, we assume Solar abundance and that the gas is at the temperature corresponding to the peak of the emissivity for that line. In systems where we have independent constraints on the gas pressure (from e.g. X-ray observations) we combine this information with the emission measure constraints to infer upper limits on the effective path length $l$ and the free electron column density $N_e$ of the hot gas as well, using equations 1 and 2. *For the M87 filament, we report both a 90\% upper limit based on a night-only spectrum as well a 2.2$\sigma$ detection of the line which includes observations taken during orbital daytime. For the latter values, 1$\sigma$ uncertainties are also listed.} 
\end{minipage}
\end{table*}  

Assuming we are at the the peak of the emissivity for each line, the emission measure $\int n_e n_p dl$ is the fundamental parameter and is linearly proportional to flux, so deeper observations will improve these limits proportionally to the square root of the exposure time. For many of the objects discussed in this section, such as Phoenix and HCG 92, the integration time would have to increase by factors of several in order to obtain emission measure limits as good as some of the other targets in our sample; a better approach would be to focus on [\ion{Fe}{xxi}] for these targets since it has a higher emissivity than the other lines. 

On the other hand, the derived column density and effective path length also depend on the pressure of the hot gas (see eqs. 1 and 2). The latter quantity in particular depends on the inverse square of the pressure, so even weak constraints on the emission measure can translate to tight constraints on the effective path length if the ambient pressure is high (i.e. in the cores of massive galaxy clusters). 

Of course, the best target in this list is the filament region in M87, where we have a tentative $2\sigma$ detection of [\ion{Fe}{xxi}]. In the next section, we show that  modest follow-up observations should readily improve these statistics and allow us to measure line kinematics. We also consider the nuclei of M87 and NGC 4696 as well.

\section{Updated HST Pipeline}
After the bulk of this analysis had been completed, the new HST Spectroscopic Legacy Archive\footnote{\url{http://archive.stsci.edu/hst/spectral_legacy/}} was released, which offers science-grade co-added spectra for galaxies and galaxy clusters observed with COS. In general these combined spectra agree well with the co-added spectra produced in this paper, which are based on the \verb"calcos" pipeline. However, the HSLA uses a different (and less conservative) method of estimating the errors, resulting in significantly smaller flux uncertainties than the errors produced by the \verb"calcos" pipeline. We therefore repeated the searches for forbidden Iron lines for each spectrum using these new HSLA spectra, and we report the results in this section.

For the filament region in M87, the HSLA full spectrum yields a positive residual at the expected location of [\ion{Fe}{xxi}] with $v_0 = -560\pm40$ km/s, $\sigma = 380^{+70}_{-60}$ km/s, and an integrated flux of  $9.6\pm1.8 \times10^{-17}$ erg s$^{-1}$ cm$^{-2}$. This is generally similar to the best-fit values in section 2, although the velocity is blueshifted by 300 km/s compared to the previous fit; this likely reflects uncertainty in continuum fitting due to the low S/N of the spectrum. The formal significance of the detection is now $5.3\sigma$ (compared to 2.2$\sigma$ from our estimate in section 2), although this should be reduced somewhat due to uncertainties in the continuum. 

We also estimated the line properties in the night-only spectrum using the HSLA error estimate. The night-only spectrum is not provided by the HSLA but we were able to reproduce their algorithm for computing the errors, which we then applied to our co-added night-only spectrum. We find $v_0 = -440^{+210}_{-120}$ km/s, $\sigma = 390^{+80}_{-110}$ km/s, and an integrated flux of  $5.6^{+2.8}_{-2.6} \times10^{-17}$ erg s$^{-1}$ cm$^{-2}$. These values are again reasonably consistent with the values from our previous analysis, albeit with higher total significance for the line (S/N = $2.2$ for this analysis, as compared to $1.0$ in section 2). 

Under the new error estimate, the [\ion{Fe}{xix}] line in the M87 filament has $v_0 = -130^{+230}_{-290}$ km/s, $\sigma = 350^{+110}_{-120}$ km/s, and an integrated flux of $F = 1.8^{+2.2}_{-2.4} \times10^{-17}$  erg s$^{-1}$ cm$^{-2}$. Again, these values are consistent with our previous estimates, but the S/N has increased from 0.3 to 0.8. 

For the nucleus of M87, the feature at the expected location of [\ion{Fe}{xxi}] now has $v_0 = 0^{+120}_{-130}$ km/s, $\sigma = 270^{+160}_{-170}$ km/s, and an integrated flux of $1.04\pm1.44 \times10^{-16}$ erg s$^{-1}$ cm$^{-2}$. These parameters are very similar to their previous best-fit values; the uncertainties are slightly decreased but the significance of the detection remains constant at 0.7$\sigma$.

For NGC 4696, we find for [\ion{Fe}{xxi}] $v_0 = 0\pm130$ km/s, $\sigma=280^{+160}_{-230}$ km/s, and $F = 0.8^{+8.8}_{-9.2} \times10^{-17}$ erg s$^{-1}$ cm$^{-2}$. This is consistent with our previous result for the line, but the S/N has actually decreased from 0.3 to 0.1. The 90\% upper limit on the flux is $1.28\times10^{-16}$ erg s$^{-1}$ cm$^{-2}$, which is slightly weaker than the previous constraint.

The upper limits in section 6 remain upper limits, but in general these limits are tighter using the HSLA spectra. The significance of the positive residual seen towards the filament in A1795 increases from 0.7$\sigma$ to 0.8$\sigma$, and the 90\% upper limit on the integrated flux decreases to $7.4\times10^{-17} \times10^{-17}$ erg s$^{-1}$ cm$^{-2}$. In the``horseshoe'' filament of Perseus, our 90\% upper limit also tightens, to $2.6\times10^{-17}$ erg s$^{-1}$ cm$^{-2}$. In the Phoenix cluster, our limits on [\ion{Fe}{xviii}] and [\ion{Fe}{xxiii}] become $1.2\times10^{-16}$ erg s$^{-1}$ cm$^{-2}$ and $1.3\times10^{-16}$ erg s$^{-1}$ cm$^{-2}$ respectively, which are moderate improvements over the values in section 6. For ZwCl 3146, the 90\% upper limits on [\ion{Fe}{xxi}], [\ion{Fe}{xix}], and [\ion{Fe}{xxiii}] respectively are now $1.1\times10^{-16}$ erg s$^{-1}$ cm$^{-2}$, $0.9\times10^{-16}$ erg s$^{-1}$ cm$^{-2}$, and $0.8\times10^{-16}$ erg s$^{-1}$ cm$^{-2}$. The first two values are moderate improvements over the limits in section 6, while the third is unchanged. Finally, in HCG 92, the HSLP data do not include the G130M data, so we are unable to update the limits for this sightline using the new error method. 

In general, using the new HSLP method of calculating errors yields similar results to the \verb"calcos" pipeline errors, but with higher inferred significances for detections and tighter constraints on non-detections. To remain conservative we report results based on the latter errors in this work, but it is straightforward to replace the values in Table 2 and elsewhere with the results based on the HSLP errors if desired.

\section{Detectability in Future Observations}

The required integration time depends on the width of the line, and we consider four example cases for the velocity dispersion: $\sigma = 50$ km/s (the thermal velocity of Iron ions), $\sigma = 150$ km/s (30\% of the sound speed for $10^7$ K gas), $\sigma = 500$ km/s (the sound speed for $10^7$ K gas), and $\sigma$ is equal to the velocity dispersion of the stars in the galaxy (roughly 450 km/s for M87 and roughly 250 km/s for NGC 4696). This yields a FWHM for [\ion{Fe}{xxi}] of 0.53\AA, 1.59\AA, 5.31\AA, and 4.78/2.66\AA{} (for M87/NGC4696) respectively for the four fiducial values of $\sigma$.

 Using our estimated flux of $7.8\times10^{-17}$ erg s$^{-1}$ cm$^{-2}$ for [\ion{Fe}{xxi}] from the filament in M87, the times required to achieve an integrated S/N of 5 would be about 6 ks, 12 ks, 33 ks, and 30 ks respectively, or approximately 3, 6, 11, and 10 orbits of integration. This is comparable to the existing integration time with the G140L grating, but the medium-resolution gratings are more sensitive and the spatial broadening will be less severe with these gratings as well. 

In the M87 nucleus, our predicted flux is similar ($7\times10^{-17}$ erg s$^{-1}$ cm$^{-2}$ for Solar metallicity; see Section 3). However, the high continuum in this region introduces noise as well. The continuum is about $1.3\times10^{-15}$ erg s$^{-1}$ cm$^{-2}$ \AA$^{-1}$, and it dominates the noise budget. In order to detect the line with an integrated S/N of 5, for the four fiducial line widths we need S/N values of about 17, 29, 54, and 48 per resel for the continuum. This requires about 40 ks, 110 ks, 390 ks, and 310 ks respectively. The latter three integration times can only be obtained through a Large GO program. It would probably be much easier to observe a sightline a few arcseconds away from the center of M87, where the continuum is much lower. If the line flux remained constant but there were no continuum, the required integration time would be reduced by a factor of 10. 

In the NGC 4696 nucleus, the continuum is less of an issue, but the line flux is very low. If we use the best-fit integrated flux of $2\times10^{-17}$ erg s$^{-1}$ cm$^{-2}$ for [\ion{Fe}{xxi}] in this sightline, then the required integration times to achieve a S/N = 5 are 50 ks, 130 ks, 410 ks, and 220 ks respectively.

The velocity resolution per resel is around 15 km/s for G130M, and this is also about the $1\sigma$ absolute calibration uncertainty for the instrument. However, depending on the width of the line, some binning may be necessary to get a S/N of 1 per bin if the integrated S/N is 5. Accounting for this binning, for our four fiducial velocity dispersions, we expect velocity resolution of about 20, 30, 60, and 40/60 km/s respectively. 

If the emission is extended, then this velocity resolution will be degraded further by spatial broadening. In section 1.1 we estimated this effect to be about 81 km/s for uniformly extended emission. Extended emission can be identified from the 2D cross-dispersion axis versus dispersion axis image, and deconvolution can be attempted if necessary, but a limiting resolution of 81 km/s is still just 15\% of the sound speed in $10^7$ K gas. A measurement of turbulent motion with $\sim$80 km/s resolution would be better than any previous measurements.

It is worth emphasizing that FUV spectrographs have much better sensitivity than their X-ray counterparts, which enables the study of these faint lines. For example, the Reflection Grating Spectrometer on XMM-Newton has an effective area of up to $\sim$ 150 cm$^2$ and the Soft X-ray Spectrometer on Hitomi had $\sim 200$ cm$^2$, while COS has more than 2000 cm$^2$ with the medium-resolution gratings. Proposed future UV spacecraft such as the World Space Observatory \citep{Shustov2009} or the Public Telescope\footnote{\url{http://www.publictelescope.org/}} will retain or improve upon the spectral resolution of COS, but these missions do lack the large collecting area of HST.

Finally, as we mentioned in the introduction, these lines can be observed in absorption as well. The absorption cross-sections are low, typically of order $\sigma_T$, but with absorption-line observations much more distant sources can be analyzed as long as they have a bright background quasar projected behind them. For [\ion{Fe}{xxi}], the optical depth is 1\% for an electron column of $8\times10^{21}$ cm$^{-2}$ (assuming Solar metallicity and a temperature of $10^{7.05}$ K). This can be achieved with an electron density of 0.2 cm$^{-3}$ over a path length of 10 kpc, which is obtainable in the center of some galaxy clusters, for example.

\section{Conclusions}

In this paper we searched for [\ion{Fe}{xxi}] $\lambda$1354.1 and [\ion{Fe}{xix}] $\lambda$1118.1 emission in publicly available COS FUV spectra towards M87 and NGC 4696. These spectra are very complex and we do not attempt to model them in full, instead focusing our analysis on the region around [\ion{Fe}{xxi}] and [\ion{Fe}{xix}]. Towards a filament in M87, we report a tentative (2.2$\sigma$) detection of [\ion{Fe}{xxi}] with an integrated flux of $7.8\pm3.6\times10^{-17}$ erg s$^{-1}$ cm$^{-2}$. This line is located very close to a nearby \ion{O}{i} airglow line, and while this airglow line seems to be negligible in these observations, we also report a 90\% upper limit based only on the data taken during orbital night. We also report 90\% upper limits on [\ion{Fe}{xxi}] towards the nuclei of M87 and NGC 4696, with values of $4.5\times10^{-16}$ erg s$^{-1}$ cm$^{-2}$, and $2.2\times10^{-16}$ erg s$^{-1}$ cm$^{-2}$ respectively, corresponding to limits on the line luminosities of $1.4\times10^{37}$ erg s$^{-1}$, and $5.1\times10^{37}$ erg s$^{-1}$.

We perform emission measure analysis on each pencil-beam column as well, inferring constraints on the electron column and the characteristic path length of the hot gas, assuming a temperature of $10^{7.05}$ K (the temperature at which [\ion{Fe}{xxi}] emissivity peaks). For the filament sightline, we infer a hot gas electron column of $9\pm4 \times10^{20}$ cm$^{-2}$ and an effective column length of $1.5\pm0.7$ kpc. For the nuclei of M87 and NGC 4696, our respective 90\% upper limits on the column density are $7.4\times10^{20}$ cm$^{-2}$ and $4.0\times10^{21}$ cm$^{-2}$ and our respective 90\% upper limits on the effective path length are 8.2 kpc and 14.1 kpc.

In both the filament sightline and the nuclear sightline we are able to compare our constraints to X-ray measurements of the emission measure of 1 keV gas by \citet{Werner2010} and \citet{Werner2013}.  In both cases our UV constraints are fully consistent with the X-ray measurements. 
 
We also examine several other sightlines towards galaxy clusters. These additional sightlines probe multiphase media, but there are not clear predictions of massive reservoirs of $10^7$ K gas along these sightlines. We do not detect any of these FUV lines in these spectra, although some of our upper limits are competitive with the M87 and NGC 4696 spectra, or with existing X-ray measurements. Our limits can be used to constrain the differential emission measure of the multiphase plasma in some cases.

Follow-up COS observations of moderate duration should be able to improve the situation dramatically, as we showed in Section 8. In particular, it is especially important to obtain medium-resolution spectra of the M87 filament, in order to confirm the weak [\ion{Fe}{xxi}] feature we identified and to measure the shape of the line. 
 
With future instrumentation, observations targeting [\ion{Fe}{xxi}] will be able to expand to additional sightlines. Multiple sightlines for a single object could be extremely valuable for mapping out the velocity structure of the intracluster and intragroup media, as well as quantifying the density fluctuations (\citealt{Churazov2012}, \citealt{Zhuravleva2014b}). Additionally, constraining the other FUV lines in Figure 1 along with [\ion{Fe}{xxi}] would yield precise constraints on the distribution of temperatures in the hot gaseous halo.

\section{Acknowledgements}
The authors would like to thank Charles Danforth and John Stocke for very helpful discussions and for assistance with reducing and co-adding COS data and the anonymous referee for thoughtful reports which helped to improve this work. Based on observations made with the NASA/ESA Hubble Space Telescope, and obtained from the Hubble Legacy Archive, which is a collaboration between the Space Telescope Science Institute (STScI/NASA), the Space Telescope European Coordinating Facility (ST-ECF/ESA) and the Canadian Astronomy Data Centre (CADC/NRC/CSA). STScI is operated by the association of Universities for Research in Astronomy, Inc. under the NASA contract NAS 5-26555.

%\bibliographystyle{mn2e}
%\bibliography{lib}
%\end{document}

\appendix

\section{Filtering \ion{O}{i} 1356 airglow emission from the G140L spectra}

In this section we discuss the geocoronal \ion{O}{i} 1356 doublet, and the efforts which were undertaken to rule out emission from this doublet in our G140L observation. At the redshift of M87, [\ion{Fe}{xxi}] has an expected wavelength of 1359.9\AA, which is just a few resels away from geocoronal \ion{O}{i} $\lambda$1358.51\AA. This is the weaker of the two lines in the \ion{O}{i} $\lambda$1356 doublet; the stronger line has $\lambda=1355.60$ \AA. \ion{O}{i} $\lambda$1356 dayglow emission is powered by collisional excitation from photoelectrons, while the nightglow is dominated by radiative recombination; the time-averaged dayglow emission looking towards nadir at 600 km (similar the orbital elevation of HST, which is around 550 km) is more than an order of magnitude brighter than the nightglow \citep{Meier1991}. We use the \verb"timefilter" routine provided in costools v. 1.0.6 to filter the data, keying on the altitude of the Sun in order to remove the data taken during local daytime (\verb"SUN_ALT" $ > -10$), which eliminates the dayglow emission. 

In order to be confident that we have accounted for all the geocoronal \ion{O}{i}$\lambda1356$, we briefly discuss the nature of the  \ion{O}{i}$\lambda1356$ nightglow emission. Most of the \ion{O}{i}$\lambda1356$ nightglow is produced at lower altitudes than the orbit of HST, and so the emission from this doublet looking towards zenith is generally negligible \citep{Syphers2012}). As the altitude increases, the zenith-looking ratio of \ion{O}{i} $\lambda1356$ to  \ion{O}{i} $\lambda1304$ gets lower and lower, since the latter have oscillator strengths which are orders of magnitude higher than the former. Some \ion{O}{i} $\lambda1304$ emission produced below the orbit of HST can therefore scatter into the line of sight, while most nightglow \ion{O}{i} $\lambda1356$ emission does not reach the detector (\citealt{Gerard1977}, \citealt{Chakrabarti1984}). This explains the relative nightglow line strengths listed in the Cycle 23 COS instrument handbook (table 7.4), in which \ion{O}{i} $\lambda1304$ is an order of magnitude brighter than  \ion{O}{i} $\lambda1356$ at orbital night.

In Figure 2, the \ion{O}{i} $\lambda1304$ triplet has nearly disappeared, due to the removal of the dayglow emission. The emission from the \ion{O}{i} $\lambda1356$ should be even weaker than the tiny residual at 1304\AA, probably by an order of magnitude or more. Consequently, we can feel confident that the spectrum in Figure 2 is essentially clean of any measurable  \ion{O}{i}$\lambda1356$ geocoronal emission.

The analysis of dataset LC7T02010 was more complicated, so we do not include results from this observation in the merged spectrum in Figure 2. Three of the ten observations within this dataset (lc7t02h0q, lc7t02hiq, and lc7t02hmq) show anomalously high \ion{O}{i} $\lambda$1354 airglow emission, which remains strong during orbital night even while \ion{O}{i} 1304 decreases. This behavior has been observed by other missions as well (e.g. \citealt{DeMajistre2005}) during periods of high geomagnetic activity. We tested the effect of combining the other seven observations in dataset LC7T02010 with the observations in Figure 2, but the restriction to orbital night removes most of the data, so that only 883 seconds of integration time remain to be added to the existing 7858 s of night-only data. The results are therefore nearly identical to those shown in Figure 2.

\end{document}